\documentclass[11pt]{article}

\usepackage{mathtools}

\usepackage{setspace}
\usepackage{lipsum}
 \usepackage{multirow}
\usepackage{fullpage}
\usepackage{amsmath}
\usepackage{amssymb}
\usepackage{setspace}
\usepackage{bbm}
\usepackage{dsfont}
\usepackage{graphics}
\usepackage{longtable}
\usepackage[font=footnotesize,labelfont=bf,justification=centerlast,width=.9\textwidth]{caption}
\usepackage{color}
\usepackage{etoolbox}

\usepackage{array}
\usepackage{hyperref}

\hypersetup{
    bookmarks=true,%
    colorlinks,%
    citecolor=blue,%
    filecolor=black,%
    linkcolor=black,%
    urlcolor=blue
}

\newcommand{\twobytwo}[4]{\left(\begin{array}{cc}#1&#2\\#3&#4\end{array}\right)}
\newcommand\rref[1]{(\ref{#1})}
\newcommand{\be}{\begin{equation}}
\newcommand{\ee}{\end{equation}}
\newcommand{\bes}{\begin{equation*}}
\newcommand{\ees}{\end{equation*}}
\newcommand{\bea}{\begin{eqnarray}}
\newcommand{\eea}{\end{eqnarray}}
\newcommand{\beas}{\begin{eqnarray*}}
\newcommand{\eeas}{\end{eqnarray*}}

\newcommand{\bmat}{\begin{bmatrix}}
\newcommand{\emat}{\end{bmatrix}}

\def\le{\left}
\def\ri{\right}

\def\Qc{{\cal Q}}

\def\z{\mathfrak{z}}

\newcommand{\CC}{\mathbb{C}} 
\newcommand{\RR}{\mathbb{R}}
\newcommand{\QQ}{\mathbb{Q}} 
 
\newcommand{\ZZ}{\mathbb{Z}}

\newcommand{\Tr}{{\rm {Tr}}}

\renewcommand{\H}{\mathbb{H}}
\newcommand{\Z}{\mathbb{Z}}

\newcommand{\ie}{{\it i.e.~}}
\def\eg{{\it e.g.~}}

\newcommand{\xOne}{E_4}
\newcommand{\xTwo}{E_6}
\newcommand{\xThree}{{\phi}_{-2,1}}
\newcommand{\xFour}{{\phi}_{0,1}}
\newcommand{\xFive}{{\phi}_{-1,2}}

\begin{document}
\numberwithin{equation}{section}
{
\begin{titlepage}
\begin{center}

\hfill \\
\hfill \\
\vskip 0.75in

{\Large \bf  Siegel Modular Forms and Black Hole Entropy}\\

\vskip 0.4in

{\large Alexandre Belin${}^a$, Alejandra Castro${}^a$, Jo\~{a}o Gomes${}^{a,b}$, and Christoph A.~Keller${}^c$}\\
\vskip 0.3in

${}^{a}${\it Institute for Theoretical Physics, University of Amsterdam,
Science Park 904, Postbus 94485, 1090 GL Amsterdam, The Netherlands} \vskip .5mm
${}^{b}${\it Institute for Theoretical Physics, University of Utrecht, Leuvenlaan 3584 CE Utrecht, The Netherlands} \vskip .5mm
${}^{c}${\it Department of Mathematics, ETH Zurich, CH-8092 Zurich, Switzerland} \vskip .5mm

\texttt{a.m.f.belin@uva.nl, a.castro@uva.nl, j.m.vieiragomes@uva.nl, christoph.keller@math.ethz.ch}

\end{center}

\vskip 0.35in

\begin{center} {\bf ABSTRACT } \end{center}
We discuss the application of Siegel Modular Forms
to Black Hole entropy counting. The role of the
Igusa cusp form $\chi_{10}$ in the D1D5P system is 
well-known, and its transformation properties
are what allows precision microstate
counting in this case. We apply a similar method to extract the Fourier coefficients of other Siegel modular and paramodular
forms, and we show that they could serve as candidates for other types
of black holes. We investigate the growth of 
their coefficients, identifying
the dominant contributions and the leading logarithmic
corrections in various regimes. We
also discuss similarities and differences
to the behavior of $\chi_{10}$,
and possible physical interpretations of such forms both from a microscopic and gravitational point of view.
\vfill

\noindent \today

\end{titlepage}
}

\newpage

\tableofcontents

\section{Introduction}

In the language of statistical physics, an extremal black hole is a zero temperature system with a huge amount of residual entropy. Understanding which features of a quantum system can account for a large degeneracy of ground states will not only unveil interesting properties of quantum gravity, but will also  uncover novel quantum systems. Our aim here is to present statistical systems, or more precisely counting formulas, that have the potential to account for the entropy of an extremal black hole. 

Our inspiration arises from supersymmetric black holes in string theory, with the most famous example being the D1D5P system first considered by Strominger \& Vafa \cite{Strominger:1996sh}. This is a situation where there has been remarkable success in accounting for the entropy of  black holes in string theory not only at leading order, but also various classes of subleading corrections. Despite the specificity of the system, there is more than one lesson to draw from this example. The lesson we want to emphasis here is the following: the function that naturally captures the microstates is a \emph{Siegel Modular Form} (SMF). 

From a physics perspective, SMFs can be seen as a class of generating functions for families of CFT$_2$ with increasing central charge. Similar to a grand canonical partition function, in addition to having a Boltzmann factor associated to, for instance, the energy, they also have a fugacity associated to the central charge of the theories. They usually appear as generating functions of supersymmetric (BPS) states, such as in \cite{Dijkgraaf:1996it,Shih:2005uc,David:2006yn} among many other examples,  for reasons that will become clearer later on. What is powerful about these types of generating functions is the mathematical structure that underlies them. The symmetry group of a SMF is not just the ordinary modular group  $SL(2,\Z)$, but the larger Siegel modular group $Sp(4,\Z)$. This is the key feature that allows us to find that the degeneracy of states is exponentially large for a wide range of parameters even when the temperature of the system is zero.  

In the following we would like to give an overview of both sides of this problem. On the macroscopic side, black holes do have very robust features which any microscopic proposal should account for. This robustness in gravity is what we would like to translate into data of the quantum system. On the microscopic side, which is the main emphasis of this work, we want to illustrate not only how one can construct generating functions with the desired features, but also present a procedure to extract the entropy systematically.

\subsection{The black hole side of the problem}

An important open question is to describe the entropy of a black hole, $S_{\rm BH}$, in terms of a suitable microscopic degeneracy, $d(\Qc)$, i.e.
\be\label{eq:BHmicro1}
S_{\rm BH}=\ln d(\Qc)~.
\ee
This equality can be made rather precise for a class of supersymmetric black holes.\footnote{For non-extremal (finite temperature) black holes the identification in \eqref{eq:BHmicro1} is much more delicate in a full quantum theory due to Hawking radiation, among other effects. For extremal but not supersymmetric black holes, there might be an analogous definition but this will depend on the details of the solution and the theory; it is not clear that an extremal black hole is generically well defined in the full quantum theory or if it is an emergent IR state.}  In particular there is compelling evidence that there is a reasonable definition of $S_{\rm BH}$ after local and non-local corrections are taken into account: in string theory, this includes both $\alpha'$ and $g_s$ corrections. Our focus will be in building candidates for $d(\Qc)$ with the guidance of universal features encoded in $S_{\rm BH}$. In the following we will review such features and how they can constraint potential candidates for $d(\Qc)$. 

 The most systematic procedure to evaluate $S_{\rm BH}$ for extremal black holes is given by the quantum entropy function,
 which was first introduced  in \cite{Sen:2008vm,Sen:2009vz}. Comprehensive reviews are given in \cite{Sen:2007qy,Mandal:2010cj}. In a nutshell, the quantum entropy function is defined via a path integral 
 \be
 Z(\Qc)_{\rm AdS_2}=\int D\phi \,e^{-S_{\rm grav}}~.
 \ee
Here the subscript ``AdS$_2$'' indicates that the path integral is performed on the near horizon geometry with suitable boundary conditions that allow for single centered black holes. The path integral is over all fields (massless and massive), and $S_{\rm grav}$ is an effective action containing both boundary terms in addition to all interactions of these fields. We use $\Qc$ as a shorthand to denote the electric and magnetic charges carried by the black hole.  

From here the entropy of the black hole is defined as follows. First,  in general $S_{\rm grav}$ will have a divergent piece due to infinite volume effects of AdS$_2$. To regulate this divergence we introduce a cutoff $L$ and eventually take $L\to \infty$. One of key the observations in \cite{Sen:2008vm,Sen:2009vz} is that from general principles of AdS$_2$/CFT$_1$, $ Z(\Qc)_{\rm AdS_2}$ can be interpreted as the partition function of a dual quantum mechanical system sitting at the boundary of AdS$_2$ on a Euclidean circle of length $L$. This allows us to interpret
\be
 Z(\Qc)_{\rm AdS_2} = {\rm Tr}_{\rm CFT_1}(e^{-L H})\, \xrightarrow[L\to \infty]{}\,  d(\Qc) e^{-E_0 L}~,
\ee
where $E_0$ is the ground state energy and $L$ is the length of the boundary circle in AdS$_2$; the infrared limit is $L\to \infty$. The macroscopic entropy is hence given by\footnote{ $ Z(\Qc)_{\rm AdS_2}$ can also contain contributions from exponentially suppressed geometries which we are ignoring. Strictly speaking, we should write in \eqref{eq:intsbh} the symbol $\approx$ where we mean that we are only extracting the contribution from the largest exponential contribution. }
\be\label{eq:intsbh}
S_{\rm BH} = \ln d(\Qc) = \lim_{L\to \infty} \le(1-L{d\over dL}\ri)\ln  Z(\Qc)_{\rm AdS_2} ~.
\ee
We stress that this relation is derived from general principles of AdS$_2$/CFT$_1$, which are carefully discussed in the above references. The  strength of this method lies in the fact that it can capture the local contributions due to the Wald entropy and non-local quantum corrections.  

We are interested in a regime where $S_{\rm BH}$ is governed by the two derivative theory of gravity for which the black hole is a smooth solution. More concretely, we want a regime where the near horizon geometry of the black hole is weakly curved. It is well known that the contribution to $S_{\rm BH}$ from the two derivative theory is proportional to the area of the horizon ($A_H$). This two derivative action also predicts the leading quantum logarithmic corrections controlled by $A_H$ in Planck units. These are the contributions to $ Z(\Qc)_{\rm AdS_2}$ that arise from the one-loop effective action of {\it all} massless fields in the low energy theory. This includes local and non-local contributions at the one-loop level.

Logarithmic corrections to black hole entropy are very powerful: they are governed by low energy data  that probe non-trivially any theory of quantum gravity that attempts to account for the black hole microstates. As such, they are a successful and robust test in several situations \cite{BanerjeeGuptaMandalEtAl2011,Sen2012b,Sen:2012cj,Sen2014}.  For CHL models both $d(\Qc)$ and its 4D(5D) supersymmetric black hole counterpart
are known explicitly, and the agreement is remarkable (see Appendix \ref{app:BH} for a quick review of this class of black holes). Logarithmic corrections have been computed as well for several other supersymmetric  configurations  \cite{Chowdhury:2014lza,Gupta:2014hxa} and using novel techniques in \cite{Keeler:2014bra,LarsenLisbao2015,KeelerLisbaoNg2016}.  
There are also very interesting results for non-extremal black holes \cite{Sen:2012dw,Charles:2015eha}.  Many of these examples do not have a microscopic counterpart yet, but their logarithmic corrections will give key clues to building a microscopic description. Understanding the statistical nature of these corrections gives a powerful insight in the quantum nature of the black hole.

To summarise, a two derivative theory of gravity predicts that
\be\label{eq:BHfull}
S_{\rm BH} = {A_H \over 4G} + w \ln {A_H \over 4G}+ \cdots~, \qquad {\rm for} ~ ~~{A_H\over 4G}\gg1 ~,
\ee
where $w$ is some numerical coefficient that depends mostly on the number of massless modes in the spectrum, among other features of the solution. Our emphasis will be in building new examples $d(\Qc)$ whose asymptotic growth has exactly this form. We approach the problem from the mathematical side, and will not be able to give a description of the matching black hole. But if a match can be found, our results give a statistical interpretation to not just the leading area law term but also $w$.

\subsection{The microscopic side of the problem}\label{sec:msp}

On the microscopic side we know that for many black holes the
entropy formula $S_{\rm BH}$ can be accounted for by modular invariance.
If the partition function in question is given by
some modular function or a Jacobi type form, we can use
$SL(2,\Z)$ invariance to obtain its asymptotic growth. 
If it has non-vanishing polar part, then we will get 
Cardy growth \cite{cardyformula}, \ie
\be\label{Cardy}
d(E) \sim e^{2\pi\sqrt{{c\over 6}E}}\ .
\ee
This is exactly the right behavior to give the Bekenstein-Hawking
entropy $S_{\rm BH}$ \cite{Strominger:1996sh}. Note however that (\ref{Cardy}) only holds in
the regime where $E \gg c$. In the language of gravity this
means that the black hole has to be very heavy. In general
we would however expect $S_{\rm BH}$ to also hold for small black holes.
The natural supergravity regime is $c \gg 1$, but not
necessarily $E \gg c$. An arbitrary Jacobi form will usually not obey
(\ref{Cardy}) in this regime
 --- that is, it will not have an extended Cardy
regime. If we want to find forms that can count the entropy of black
holes, we therefore need quite special forms, or, more precisely,
families of forms. 

The best known example of this is the Siegel modular form
$\chi_{10}$, or rather its reciprocal, which is the generating function appearing in the counting of 1/4-BPS dyons in four dimensions \cite{Dijkgraaf:1996it}. We interpret it as the
generating function of a family of Jacobi forms. Its symmetries are not simply the expected
$SL(2,\Z)$, but are enhanced to the Siegel modular group $Sp(4,\Z)$.
Its Fourier coefficients $d(E)$ then have an extended Cardy regime, which allows
it to be interpreted as describing the entropy of the D1D5P black hole in four dimensions.
It is very natural to suspect that the extended Cardy regime
and the enhanced symmetry group should be related; see also \cite{LVM}. This 
motivates the study of other SMFs.

Our general strategy is therefore to investigate the
space of SMFs and their generalizations,
so-called Siegel paramodular forms.\footnote{To avoid cluttering, we will refer to both Siegel modular forms and Siegel paramodular forms as SMFs.} In the D1D5P case
$Sp(4,\Z)$ transformations  is one of the key symmetries that allows one to compute the 
dominant contribution to the entropy and its logarithmic
corrections. We apply the same strategy to more general
SMFs. Our goal is to work out the final result in terms
of only a few properties of the underlying form. In fact
we find that the result depends only on the weight of
the SMF, the position of its poles, and some properties
of the residue at the poles. 

As we sample the space of SMFs, we will be interested in identifying SMFs that have the correct features to account for black hole entropy. Following the criteria discussed in \cite{Hartman:2014oaa,Haehl:2014yla,Belin:2014fna}, a necessary feature we will require to make this identification is:
\begin{quote}
\emph{There is an extended Cardy regime: the exponential growth in \eqref{Cardy} is valid even if $E\sim c\gg1$. This is the natural scale in supergravity where we expect black holes to dominate the ensemble.} 
\end{quote}
We will identify several SMFs that satisfy this condition. And within this class it is important to make two further distinctions:\footnote{Demanding that the Cardy regime extends for $c\gg E\gg1$ is a much stronger condition than that demanded in \cite{Hartman:2014oaa}.  Therefore, their notion of sparseness (which allows Hagedorn growth) is much looser than ours.}
\begin{enumerate}
\item The Cardy regime  extends also to $c\gg E\gg1$. Moreover, the  perturbative part of the spectrum, {\it i.e.} polar states, does not exhibit Hagedorn growth. This corresponds to a very sparse low energy spectrum, which hints that there is a supergravity regime. 

\item The Cardy regime breaks down for $c\gg E\gg1$, and a Hagedorn spectrum takes its place. This type of behavior is more compatible with a string theory spectrum (with no semi-classical supergravity regime). 
\end{enumerate}
We will elaborate more on these conditions as we go along with our analysis. Having an extended Cardy regime should be viewed as necessary for the SMFs to have a black hole (or gravitational) interpretation, but it might not be sufficient. Moreover, satisfying the first condition is necessary for there to be a supergravity description of the black hole. In our opinion, satisfying this requirement is a very compelling reason to study these cases further.  And an important part of our results is that we can meet the first condition for examples that deviate significantly from the well known case of  $\chi_{10}$ and cousins examples.   

In some cases we can find physical interpretations
of the Siegel modular forms we study. For
instance, $1/\chi_{10}$  is, up to an overall factor, the generating function 
of the symmetric orbifold of $K3$. Similar forms
exist for the symmetric orbifolds of higher dimensional
Calabi-Yau manifolds. The ultimate goal would of course
be to identify the CFT and gravity (string) dual of those forms.
We discuss some steps in that direction.


\section{Siegel Modular Forms}\label{sec:smf}

Our starting point is to consider generating functions which are of the form
\bea\label{eq:s10}
{\Phi}(\rho,\tau,z)= \sum_{m,n,l} d(m,n,l)p^m q^n y^l ~,
\eea
where $p=e^{2\pi i \rho}$, $q=e^{2\pi i\tau}$, and $y=e^{2\pi iz}$; for now the domain of $(m,n,l)$ is unspecified and it will be narrowed as needed.  We can alternatively write
\be
{\Phi}(\rho,\tau,z)= \sum_{m} \varphi_{k,m}(\tau,z) p^m ~,
\ee
where the Fourier coefficients of $\varphi_{k,m}$ are given by $d(m,n,l)$. We are interested in cases where $\varphi_{k,m}(\tau,z)$ is 
 a \emph{Jacobi form}, where $k$ is the \emph{weight} and $m$ is the \emph{index}. The definition of Jacobi forms and some of their properties are listed in appendix \ref{app:jf}. In addition, here we will be interested in a rather specific class of generating functions ${\Phi}$: we will also consider functions that are symmetric up to a sign  under the exchange of $p$ and $q$,
\be\label{eq:s11}
{\Phi}(\rho,\tau,z)= (-1)^k{\Phi}(\tau,\rho,z)~.
\ee
This transformation, combined with the transformation properties of Jacobi forms, generates the full Siegel modular group $Sp(4,\Z)$, so that 
${\Phi}$ has the transformation properties of a so-called \emph{Siegel modular form} (SMF). 

In the following we will review various properties of SMFs. In addition to its transformation properties with respect to $Sp(4,\Z)$, we will discuss its zeros and poles, introduce the concept of exponential lift, and present generalizations for paramodular groups. 

\subsection{Basic definitions and properties of SMFs}

In this section we will summarise the key features of SMFs we will use; for a more complete and mathematical discussion see \cite{MR2385372,9781468491647}, and for a review of SMFs in string theory see e.g. \cite{Sen:2007qy, Dabholkar:2012nd}.  We start with classical holomorphic Siegel modular forms of degree $g=2$ of the full group $Sp(4,\Z)$ of weight $k$, whose space we denote by $M_k =M_k(\Gamma_2)$.  We take\footnote{In comparison to, e.g., \cite{Sen:2007qy} we have $z=v$ and $\tau =\sigma$, and relative to  \cite{Gritsenko:1999fk}, we have $\omega=\sigma$.} 
\be\label{Omegadef}
\Omega = \left(\begin{array}{cc}\tau&z\\z&\rho\end{array}\right)\ .
\ee
The Siegel upper half plane $\H_2$ is given by
\be
\det(\Im(\Omega)) > 0 \ , \qquad \Tr (\Im(\Omega)) > 0\ .
\ee
A matrix $\gamma \in Sp(4,\Z)$ is given by
\be
\gamma=\twobytwo{A}{B}{C}{D}~,
\ee
with the $2\times2$ blocks satisfying
\be
A B^T = BA^T\ , \qquad CD^T = DC^T\ , \qquad A D^T - BC^T= \mathds{1}_2\ .
\ee
The action of $\gamma$ on $\Omega$ is given by
\be
\gamma(\Omega)= (A\Omega+B)(C\Omega+D)^{-1}\ .
\ee

A Siegel modular form ${\Phi}(\Omega)$ of weight $k$ is a holomorphic function on the Siegel upper half plane that satisfies
\be\label{eq:tz}
{\Phi}( (A\Omega+B)(C\Omega+D)^{-1})=\det(C\Omega+D)^k {\Phi}(\Omega)~.
\ee
Note that 
\be\label{eq:flip}
\twobytwo{\sigma_1}{0}{0}{\sigma_1} \in Sp(4,\Z)~,
\ee
exchanges $\rho \leftrightarrow \tau$; SMF of even weight are invariant under this transformation. 

By definition SMFs are \emph{holomorphic}. This
in particular implies that they have non-negative weight. The space of classical Siegel modular forms 
generated by just five generators, $E^{(2)}_4$, $E^{(2)}_6$, $\chi_{10}$, $\chi_{12}$, $\chi_{35}$,
whose weights are given by their subscripts \cite{Igusa35,Igusa}. 
Here $E_{4,6}^{(2)}$ are the genus 2 Eisenstein
series of weight 4 and 6.
The only relation between
those generators is that $\chi_{35}^2$ can be expressed as a polynomial
of the other four generators. The ring of Siegel modular forms
is thus
given by 
\be\label{eq:ring}
M_k = \CC[E^{(2)}_4,E^{(2)}_6,\chi_{10},\chi_{12}]\oplus
\chi_{35}\cdot\CC[E_4^{(2)},E_6^{(2)},\chi_{10},\chi_{12}]~.
\ee
We will present explicit properties of these forms in section \ref{sec:examples}. 
We say ${\Phi}$ is a \emph{cusp form} if
\be
\lim_{t\rightarrow\infty}
{\Phi}\left(\begin{array}{cc}\tau&0\\0&it\end{array}\right)=0\ ,
\ee
and denote the space of such cusp forms $S_k$.

For ${\Phi}\in M_k$, we can write down
a Fourier expansion in $p=e^{2\pi i\rho}$, i.e.
\be\label{eq:ZJM}
{\Phi}(\Omega)= \sum_{m} \varphi_{k,m}(\tau,z) p^m ~.
\ee
The coefficients
$\varphi_{k,m}$ of this expansion are then Jacobi forms of
weight $k$ and index $m$ (see section 8 of \cite{MR2385372}). 
To see this explicitly, note that 
\be\label{SLelement}
\gamma=\left(
\begin{array}{cccc}
 a & 0 & b & 0 \\
 0 & 1 & 0 & 0 \\
 c & 0 & d & 0 \\
 0 & 0 & 0 & 1 \\
\end{array}
\right)~,\quad {\rm with }~~ad-bc =1~,
\ee
gives the coordinate transformation
\be
\tau \mapsto \frac{a\tau+b}{c\tau+d}~,\qquad
z \mapsto \frac{z}{c \tau +d}~, \qquad
\rho \mapsto \rho -\frac{c z^2}{c \tau +d } ~.
\ee
This gives the correct transformation behavior
for $\varphi_{k,m}$ in \eqref{eq:jf1}. Moreover the transformation
\be\label{shiftelement}
\gamma=\left(
\begin{array}{cccc}
 1 & 0 & 0 & \mu  \\
 \lambda  & 1 & \mu  & 0 \\
 0 & 0 & 1 & -\lambda  \\
 0 & 0 & 0 & 1 \\
\end{array}
\right)~,
\ee
leads to the other transformation property for Jacobi forms in \eqref{eq:jf2}. It is interesting to note that $Sp(4,\Z)$ is generated by \eqref{eq:flip}, \eqref{SLelement}, and \eqref{shiftelement} \cite{9781468491647}: these are the basic ingredients to construct a SMF.

For our purposes, holomorphic SMF do not have the right properties.
In particular their Fourier-Jacobi coefficients $\varphi_{k,m}$ are
true Jacobi forms, whose coefficients only grow polynomially. For 
black hole entropies, we expect exponential (or, more precisely, Cardy type) growth. We will therefore also consider meromorphic SMF. In that case
the $\varphi_{k,m}$ will still have the correct Jacobi form transformation
properties, but they are no longer true Jacobi forms, but rather weak
Jacobi forms, or even meromorphic Jacobi forms, in which case the Fourier coefficients can have exponential growth (see Appendix \ref{app:jf}).
Meromorphic SMF can be obtained from rational functions of 
classical SMF.
In the physics literature the best known example for this is 
\be
\frac{1}{\chi_{10}}\ ,
\ee
the reciprocal of the Igusa cusp form $\chi_{10}$. The goal of
our paper is to go beyond this case.

\subsection{SMFs for paramodular groups}


Next we want to generalize the concept of Siegel modular forms
to so-called \emph{paramodular groups}, that is certain subgroups
of $Sp(4,\RR)$.

The paramodular group $\Gamma_N$ of level $N$  is defined as \cite{MR2208781}
\be
\Gamma_N :=\left[\begin{array}{cccc} 
	\Z & N\Z &\Z&\Z\\
	\Z &\Z&\Z&N^{-1}\Z\\
	\Z& N\Z&\Z&\Z\\
	N\Z&N\Z&N\Z&\Z
\end{array}\right] \cap Sp(4,\QQ).
\ee
We denote by $M_k(\Gamma_N)$ the space of Siegel modular
forms of weight $k$ under $\Gamma_N$.
The paramodular group has an extension
\be
\Gamma_N^+ = \Gamma_N \cup \Gamma_N V_N\ , \qquad
V_N = \frac{1}{\sqrt{N}}
\left(\begin{array}{cccc} 
	0&N&0&0\\
	1&0&0&0\\
	0&0&0&1\\
	0&0&N&0
\end{array}\right)\ .
\ee
Note that $\Gamma_N$ contains both (\ref{SLelement})
and (\ref{shiftelement}). The Fourier-Jacobi
identity of a form $\Phi \in M_k(\Gamma_N)$ thus
again leads to Jacobi forms of weight $k$ and index $m$.
Note however that $\Phi$ has to be invariant under
\be
\left(\begin{array}{cccc} 
	1&0&0&0\\
	0&1&0&N^{-1}\\
	0&0&1&0\\
	0&0&0&1
\end{array}\right)~,
\ee
which means that all non-vanishing powers of $p$ 
are multiples of $N$. It follows that we
get a family of Jacobi forms with index $Nm$
rather than just $m$ as in the original case.

\subsection{Exponential lifts}\label{sec:explift}
Through the Fourier-Jacobi expansion, we know how to obtain
Jacobi forms from SMF. Let us now discuss the converse question:
given some type of Jacobi form, can we lift it to a SMF?
It turns out that this is possible for certain forms, and that
there are in fact two types of lifts: additive and exponential lifts. Our focus will be mostly on the later; the additive lift will play a minor role around \eqref{eq:f10}.

The \emph{exponential lift} is described in Theorem 2.1 
of \cite{MR1616929}, which first portion states:
\begin{quote}
Let $\varphi\in J^{nh}_{0,t}$ be a nearly holomorphic Jacobi form of weight 0 and index $t$
with integral coefficients
\be
\varphi(\tau,z)= \sum_{n,l} c(n,l)q^n y^l\ .
\ee
Define
\be
A = \frac{1}{24}\sum_l c(0,l)~,\qquad  B = \frac{1}{2}\sum_{l>0} l c(0,l)~, \qquad C = \frac{1}{4}\sum_l l^2 c(0,l)\ .
\ee
Then the exponential lift of $\varphi$ is the product
\be\label{explift}
\textrm{Exp-Lift}(\varphi)(\Omega)= q^A y^B p^C \prod_{\substack{n,l,m\in\ZZ
		\\(n,l,m)>0}} (1-q^n y^l p^{tm})^{c(nm,l)}\ ,
\ee
where $(n,l,m)>0$ means $m >0 \lor (m=0 \land n>0) \lor (n=m=0 \land l <0)$, and it defines
a meromorphic modular form of weight $\frac{1}{2}c(0,0)$ with respect to $\Gamma^+_t$.
It has a character (or a multiplier system if the weight is half-integral) induced by $v^{24A}_\eta \times v^{2B}_H$.
Here $v_\eta$ is a 24th root of unity, and $v_H=\pm1$.
\end{quote}
Note that if $\varphi$ has a pole at $\tau =i\infty$,
\ie if it really is a nearly holomorphic form, then
the infinite product contains terms with negative $n$.
If $\varphi$ is a weak Jacobi form, then
we actually have $C=tA$.

There is an analogue statement for forms of half-integer weights.
We can use the Hecke operator $U_2$ (see appendix~\ref{app:Hecke})
which maps
any $\varphi(\tau,z)\in J_{0,d/2}$ to
$\varphi(\tau,2z) \in J_{0,2d}$. (The converse is obviously
not true.) 
For half-integer $t$, we can then apply the above theorem
to $\varphi|U_2$
to get a Siegel paramodular form in $M_0(\Gamma^+_{4t})$,
possibly with a multiplier system. Note that half-integer
index weak Jacobi forms have automatically $c(0,0)=0$, so
that their lifts have weight 0.

The exponential lift can be naturally split into two
factors, namely
\be\label{explift2}
\textrm{Exp-Lift}(\varphi)(\Omega)= q^A y^B p^C 
\prod_{\substack{(n,l)>0} }(1-q^n y^l)^{c(0,l)}\times
\prod_{\substack{n,l,m\in\ZZ
		\\m>0}} (1-q^n y^l p^{tm})^{c(nm,l)}\ .
\ee
Here $(n,l)>0$ means $n>0 \lor (n=0\land l<0)$. The second factor can be naturally written in terms
of Hecke operators $T_-(r)$, namely as
\be\label{eq:symprod}
\exp\left( - \sum_{r\geq1} r^{-1}p^{tr}\varphi|T_-(r) \right) = \frac{1}{\Phi_\varphi}\ .
\ee
If $\varphi$ is some elliptic genus or
partition function $\chi$ of a CFT,
then $\Phi_\varphi$ is the generating function
for the partition functions of the symmetric orbifolds
of that theory,
\be
\Phi_\chi= \sum_{r=0}^\infty p^{tr} \chi(\tau,z; {\rm Sym}^r(M))\ .
\ee
A famous example of this is the Igusa cusp form $\chi_{10}$, which
is the exponential lift of the weak Jacobi form $2\phi_{0,1}$.
Another example is $\chi_{35}$, which is the exponential lift
of a nearly holomorphic Jacobi form of weight 0 and index 1,
namely $\phi_{0,1}|T_2-2\phi_{0,1}$. We will return to this in
sections \ref{sec:examples} and \ref{sec:cft}.

%

%
%

\subsection{Zeros and poles}\label{sec:zp}

Let us now discuss the zeros and poles of meromorphic SMF that can be cast as exponential lifts, which is the second portion in Theorem 2.1  of \cite{MR1616929}. These 
zeros and poles are located on the divisors of the SMF, and for a SMF that has a product expansion (such as \eqref{explift}) it is rather simple to identify them: Choosing $\tau, z, \rho$ such that $q^n y^l p^{tm}=1$ in one of the factors will make that factor vanish, so that the product either vanishes or diverges. Because of the invariance under $\Gamma_t^+$, 
divisors will always come as orbits of $\Gamma_t^+$.


To describe the divisors of lifted SMF, it is useful to introduce Humbert surfaces: this is how we will package the orbits of $\Gamma_t^+$.
We are following section 1.3 of \cite{MR1616929} here.
Define $\ell = (e,a,-\frac{b}{2t},c,f)$ with
$e,a,b,c,f \in \Z$ and $\text{gcd}(e,a,b,c,f)=1$.
We define its discriminant as
\be\label{eq:DD}
D(\ell)=2t(\ell,\ell) = b^2 -4tef -4tac\ .
\ee
It turns out that 
there is a natural action of $\Gamma_t^+$ on $\ell$
that leaves $D(\ell)$ invariant. $\ell$ then
defines a divisor in $\H_2$ via the quadratic equation
\be\label{Humzero}
tf (z^2-\tau\rho) + tc\rho +bz+a\tau+e =0\ .
\ee
The crucial observation in \cite{MR1616929} is that \emph{all} zeros and poles of SMF that are exponential lifts are given by Humbert surfaces
$H_D(b)$. These divisors can always be written as 
\be\label{eq:hdb}
H_D(b) = \pi^+_t(\{Z \in \H_2: a\tau + bz + t\rho =0 \} )\ ,
\ee
where $\pi^+_t$ is the set of images of $\Gamma^+_t$.
The discriminant $D$ is given by $D= b^2- 4ta$ and $b\mod 2t $.
This determines its position, but each divisor has its on multiplicity (or degree). In general, the divisors of the exponential
lift (\ref{explift}) are given by the Humbert surfaces
\be
\sum_{D,b} m_{D,b} H_D(b)~,
\ee
and the multiplicities $m_{D,b}$ are given by
\be
m_{D,b}= \sum_{n>0} c(n^2a,nb)\ ,
\ee
where $c(n,l)$ are the Fourier coefficients of
the underlying form $\varphi$.
From this we see that the Humbert surface of maximal
discriminant $D$ comes from the term with maximal
polarity of $\varphi\in J^{nh}_{0,t}$.

In the following section an important case will be the Humbert surface
$H_1(1)$ for exponential lifts of weak Jacobi forms. Note that due to the transformation
\be
\gamma = \left(
\begin{array}{cccc}
 1 & t & 0 & 0 \\
 0 & 1 & 0 & 0 \\
 0 & 0 & 1 & 0 \\
 0 & 0 & -t & 1 \\
\end{array}
\right) \in \Gamma_t^+\ ,
\ee
which maps $z\mapsto z+t\rho$, the divisor $z=0$ is
in $H_1(1)$. The behavior near $z=0$ will be vital as we extract asymptotic formulas.  
The crucial identity here is the `Witten index' identity, i.e. 
for $\varphi \in J^{weak}_{0,t}$,
$$\sum_l c(n,l) = 0\,,\quad \forall\, n>0~.$$
The leading zero or pole near $z=0$, up to numerical coefficients, is then
\begin{align}\label{liftresidue}
&q^A p^{tA} \prod_{m>0} (1- p^{tm})^{24A} \prod_{n>0} (1-q^n)^{24A} \prod_{l<0} (1-y^l)^{c(0,l)}\cr
&\sim z^{m_{1,1}} q^A p^{tA} \prod_{m>0} (1- p^{tm})^{24A} \prod_{n>0} (1-q^n)^{24A}\cr
&= z^{m_{1,1}} \eta(\tau)^{24A} \eta(t\rho)^{24A}
\end{align}
with
\be\label{eq:m11}
m_{1,1}={\sum_{l<0} c(0,l)}~.
\ee
Note that the zero (or pole) has indeed multiplicity $m_{1,1}$. 


\subsection{Mapping to CFT variables}

To end this section, we will set our conventions on how some basic data of a SMF is mapped to CFT jargon.  In particular, we will translate quantities such as $(m,n,l)$ to the quantum numbers, anomalies and Casimir energies.

Very broadly speaking, from its definition, a SMF is intimately related to Jacobi forms via \eqref{eq:ZJM}. And a Jacobi form, as reviewed in appendix \ref{sec:JFCFT}, can be thought as a generalized partition function of a system whose algebra contains at least one copy of Virasoro and a $U(1)$ Kac-Moody current. For instance, $\varphi_{k,m}$ can be interpreted as the Elliptic genus of a SCFT or the partition function of a chiral theory. Hence, we can interpret $\varphi_{k,m}$ as
\be
{\rm Tr}_{\cal H}( q^{L_0-c/24} y^{J_0})~,\qquad {\rm or} \qquad \Tr_{RR}\le((-1)^F(-1)^{\bar F}q^{L_0-c/24}{\bar q}^{\bar L_0 -c/24}y^{J_0}\ri)~,
\ee    
where $L_0$ and $J_0$  are the zero modes of the Virasoro algebra and Kac-Moody current respectively. If we denote $E$ and $J$ the eigenvalues of $L_0$ and $J_0$, the relation to the notation used above is straight  forward:
\be
E=n~,\qquad J= l~.
\ee

The index $m$ of $\varphi$ is the level of the Kac-Moody algebra, i.e. the anomaly in the current OPE. An `effective' central charge can be inferred from the most polar term in $\varphi_{k,m}$. We recall that polar terms are those whose discriminant is negative: $\Delta= 4nm-l^2 <0$. In this sense the maximal polarity is  the analogous of  the Casimir energy of the ground state. If we denote the most polar term as $(n_0,l_0)$, then we identify schematically the an effective central charge as\footnote{Strictly speaking, here we use "effective central charge" to denote the quantity that controls the asymptotic growth of states at very high energies, i.e. the Cardy regime. Its precise relation to the central charge can be derived depending on the physical origin of $\varphi$, but for now \eqref{eq:ceff} is enough.}
\be\label{eq:ceff}
\sqrt{l_0^2-4n_0m}= {c_{\rm eff}\over 24}~. 
\ee
 For weak Jacobi forms $l_0$ is bounded by the index $m$, while for nearly holomorphic forms $(n_0,l_0)$, are arbitrary. In either case, a large `$c$' limit is closely related to a large $m$ limit, and for this reason it is useful to think of $m$ as controlling the central charge. For supersymmetric examples this can be made more sharp since $c=6m$, but this relation is not generic. 

SMFs that have the most natural interpretation as describing a family of CFTs are those  that can be cast as an exponential lift. Calling $\varphi_{k,t}$ in \eqref{explift} \emph{ the seed}, then the resulting SMF is the generating function of symmetric products of $\varphi_{k,t}$. 
The coefficient of $q^m$, that is the weak Jacobi form of index $m$, then corresponds to the symmetric orbifold or order $r:=m/t$.
If we denote by $D$ the discriminant of the most polar term in the seed, then this $r$-th symmetric orbifold has a polar term of maximal polarity
\be
Dr^2 = D\frac{m^2}{t^2}=l_0^2-4n_0 m ~,
\ee
where on the right-hand side $l_0$ and $n_0$ are the data of the most polar state in the $r$-th symmetric orbifold. 

\section{Asymptotic growth and log tails}\label{sec:asymp}

In this section we will obtain the asymptotic behavior of the Fourier coefficients of a class of meromorphic SMF. Our analysis follows very closely the results in \cite{JatkarSen2006,David:2006yn,Sen:2007qy}, which is specific to the reciprocal of the Igusa cusp form $\chi_{10}$ and its cousin functions for CHL models.  As we will show, the key is to exploit the zeros of ${\Phi}(\Omega)$: this will allow us to identify generating functions that have the desired physical properties and, moreover, we can extract the leading and subleading behavior easily.  

\subsection{Poles and growth behavior}

To start, lets estimate the leading growth behavior. For reasons that will become clear shortly, it will be useful to introduce some notation. In particular we will introduce $so(2,1)$ vectors, whose inner product is given by
\be
X\cdot Y= (X^1,X^2,X^3)\cdot (Y^1,Y^2,Y^3)= X^1Y^2 + X^2 Y^1-2X^3Y^3~.
\ee
Using this notation, the Fourier coefficients of the reciprocal of ${\Phi}(\Omega)$ are given by
\be\label{eq:sf1}
d({\cal Q})= \int_{\cal C} d\tau d\rho dz\, e^{-2\pi i \Qc\cdot Y}\, {1\over \Phi(\Omega)}~,
\ee
where, relative to \eqref{eq:s10}, $d(\Qc)\equiv d(m,n,l)$ and
\be\label{eq:Qdef}
{\cal Q}:=(m,n,l/2)~,\quad Y:=(\tau,\rho,-z)~.
\ee
The integration contour $\cal C$ in \eqref{eq:sf1} is chosen according to the domain in which we want to compute the degeneracy, though many asymptotic properties are not sensitive to the details of this choice. We will elaborate more on this as we examine our examples.

Our goal is to obtain an asymptotic formula for $d({\cal Q})$, i.e. we want to estimate the Fourier coefficients in a regime where all the entries  in $\Qc$ are large, and $\Qc^2$ is positive. We schematically write this scaling regime as
\be\label{eq:largeQ}
{\Qc}^2\gg1~.
\ee
Typically these states correspond to black hole states.
The gravitational counterpart of \eqref{eq:largeQ} is roughly $A_H/4G\gg1$, i.e. a smooth and weakly curved black hole solution. 
More importantly, we are looking for examples where the growth is exponentially large in this regime. To achieve this, we will consider  functions $\Phi(\Omega)$ that obey the following properties 
\begin{enumerate}
\item $1/\Phi(\Omega)$ is a meromorphic SMF with poles in the Siegel upper half plane. The simplest way to build such a function is by taking the reciprocal of a cusp SMF.
\item $\Phi(\Omega)$ can be cast as an \emph{exponential lift} as defined in section \ref{sec:explift}.
\end{enumerate}
Let's briefly justify our choices. If $1/\Phi(\Omega)$  is meromorphic it is rather easy to perform at least one of the integrals in \eqref{eq:sf1}: we can simply do a residue integral around the appropriate contour.\footnote{The meromorphicity of $1/\Phi(\Omega)$ highlights that $d({\cal Q})$ depends on the contour: as ${\cal C}$ crosses a pole we get a jump in $d({\cal Q})$. This is the well-known phenomenon of wall crossing, and while very interesting, we will not explore this subject. We will comment on this in section \ref{sec:discussion}.}  This simplifies greatly the integrand, since to extract $d({\cal Q})$  in principle we only need the residues of $1/\Phi(\Omega)$. Our second choice is more restrictive, but rather powerful. If $\Phi(\Omega)$ has a product expansion it is possible to read off the residues at a given pole, and moreover  to locate of all  divisors in  $\Phi(\Omega)$. This is crucial since we really don't want to keep track of every detail in $d({\cal Q})$: we want a practical algorithm to estimate the largest contribution in \eqref{eq:sf1} for large values of $\Qc$.

With these ingredients in hand we can make a first estimate of the behavior of  $d({\cal Q})$. Since we are considering exponentials lifts, the poles of $1/\Phi(\Omega)$ are given by the Humbert surfaces \eqref{Humzero}, which is nothing more complicated than a quadratic equation for $Y$. To identify the most dominant pole, we will need to add one assumption: for $\Qc^2 \gg1$  we can assume that  the integrand in \eqref{eq:sf1} is dominated by the explicit exponential factor. This in particular means that the residue of $\Phi(\Omega)$ does not compete with  $e^{-2\pi i \Qc\cdot Y}$;\footnote{Checking the validity of this assumption relies on the value of $Y$ near the saddle point. For our discussion,  we see from \eqref{eq:ex1} that as $\Qc^2 \gg1$ we roughly have $Y\sim O(1)$ and hence the residue of  $\Phi(\Omega)$ near this most dominant pole is of order one. However, this is a very heuristic argument that we will revisit as we move along.} of course this assumption has to be checked (and refined) for each example, but for now we will take it as given. Hence, in order to find the most dominant pole in the regime \eqref{eq:largeQ} we need to extremize
\be\label{eq:f-ext}
f(\lambda)= {\cal Q} \cdot Y + \lambda(-{1\over 2} t\,f\, Y^2 +\beta \cdot Y +e) ~.
\ee
Here the Lagrange multiplier $\lambda$ constrains $Y$ to be on the divisor \eqref{Humzero}. Adapting  the notation used in section \ref{sec:zp} to the  $so(2,1)$ notation used here, we have
\be\label{eq:abctf}
 \beta:=(t\,c,a,{b\over 2}) ~.
\ee
Extremizing $f(\lambda)$ gives
\be\label{eq:ex1}
\lambda = \pm i \sqrt{\frac{2{\cal Q}^2}{D}}\ , \qquad Y_{\rm max} = \frac{1}{t f}\left(\frac{\cal Q}{\lambda}+\beta\right) ~,
\ee
where
\be\label{eq:x1}
\beta^2+2t f e = -\frac12 D ~, \quad {\cal Q}^2= 2 (mn -{l^2\over 4}) ~,
\ee
and $D$ is given by \eqref{eq:DD}. After neglecting a phase,\footnote{Replacing \eqref{eq:ex1} gives $$e^{-i2\pi {{\cal Q}^2\over t f \lambda}+ 2\pi i {\Qc \cdot \beta\over tf} }~.$$ Since $\Qc \cdot \beta /(tf)$ is a rational number, its potential contribution is at most a phase.} we then get that at the extremum \eqref{eq:ex1} the leading behavior of the Fourier coefficient is
\be\label{eq:ddq}
d({\cal Q})\sim e^{-i2\pi {{\cal Q}^2\over t f \lambda}} = e^{-{\pi\over t f}\sqrt{2 D{\cal Q}^2}}~,
\ee
where we selected the minus sign in \eqref{eq:ex1} since $D>0$. The dominant contribution is thus for $f=-1$ and $D$ of maximal polarity, which leads to
\be\label{eq:z4}
d({\cal Q})\sim e^{\pi\sqrt{2 D{\cal Q}^2}/t} = e^{\pi\sqrt{(4mn-l^2)D}/t}\ .
\ee

At this stage it is useful to compare this result with the ordinary Cardy formula. Comparing \eqref{eq:z4} with \eqref{eq:c4} we have
\be\label{eq:Dfin}
{D\over t^2}=({l_0^2}-4n_0 m){1\over m^2} ~.
\ee
This is in agreement with our results in appendix \ref{app:cardy}. Recall that we are considering an exponential lift of $\varphi\in J^{nh}_{0,t}$, and hence  we should compare with the $r$-th symmetric product of $\varphi$. The relation in \eqref{eq:Dfin} shows that the Humbert surface that dominates the residue is indeed correctly related to the term with maximal polarity in a Jacobi from of degree $m=tr$. 

As in \cite{Sen:2007qy}, we can use translation symmetry to restrict further the remaining integers in \eqref{eq:abctf}. Taking $\rho\to \rho+1/t$ allows us to set $a=0$, translations $\tau\to \tau+1$ can be used to set $c=0$. Then shifts $z\to z+1$  lower $b\to b-2t$, which is compatible with $b$ mod $2t$.  And $e$  is finally fixed by $D$ and $b$ via \eqref{eq:x1}. This allows us to identify the most dominant pole as
\be\label{eq:dompole}
t(\tau\rho-z^2)+b z + e=0~.
\ee  
In many of the cases we will discuss in section \ref{sec:examples} we will be allowed to set $e=0$, $b=1$ and hence $D=1$: this corresponds to the Humbert surface $H_1(1)$. 

It is important to emphasize that in \eqref{eq:largeQ} we don't need for all entries to be equally large: there can be relative scalings among each component.   This leads to a powerful conclusion: assuming that $1/\Phi(\Omega)$ has a pole, we can obtain an asymptotic growth as in \eqref{eq:z4} for a wide range of energies relative to the central charge. More concretely, in terms of the components of $\Qc$, we have exponential growth in three general cases:
\begin{description}
\item[I. $n\gg 1$, $m\sim 1$]: This is the usual Cardy regime, where the energy of the state is much bigger than the central charge of the system, which is also proportional to the index. Recall that the index controls the maximal polarity of the Jacobi form which controls the validity of the Cardy regime. 
\item[II. $n\sim m\gg 1$]: Here energies are comparable to the central charge. Gravitational systems for which this scaling is relevant are, for example, BPS black holes in  ${\cal N}=4,8$ in four dimensional supergravity (see Appendix \ref{app:BH}). The BTZ black hole in three dimensions \cite{BanadosHenneauxTeitelboimEtAl1993} falls as well in this category.  
\item[III. $m\gg n\gg 1$]: We can naively access the opposite scaling as {\bf I} due to  the exchange symmetry among $\rho$ and $\tau$ of SMFs. Note however that this regime is not the exact opposite of regime {\bf I}, since we do not hold $n$ fixed as in the Cardy regime. The access to this regime does depend on how freely we can choose the contour ${\cal C}$ as we will see in the examples.  A gravitational system for which this regime is relevant is the 5D BMPV black hole: in the type IIB frame we have $m\sim Q_1Q_5$ and $n\sim P$; see e.g. \cite{Sen:2012cj,Castro:2008ys,Banerjee:2008ag}.  
\end{description}
For sake of simplicity, in the above classification we have omitted the scaling properties of $l$, but it can easily be incorporated. And of course more variants could be included, but these three regimes will suffice to illustrate the properties of  $d(\Qc)$. 

Even though all three cases listed above have the same leading behavior, given by \eqref{eq:z4}, the subleading corrections are sensitive to the details of the relative scalings of the components in $\Qc$. This is extremely important if we wanted to identify $d(\Qc)$ with the entropy of a gravitational system: our aim is to not just capture the leading area contributions, but account for subleading corrections. In the following we will show how to extract this information and subtleties that might arise. 

\subsection{New and old examples }\label{sec:examples}

In this section we will list a few examples of meromorphic SMF and properties of its Fourier coefficients. The first two examples mostly involve SMF built out of exponential lifts of weak Jacobi forms. The later examples are more exotic, and involve the less explored behavior of $\chi_{35}$ and $\chi_{12}$.

\subsubsection{The Igusa cusp form $\chi_{10}$}\label{sec:chi10}

The most successful example of this program is given by the counting formula that captures the degeneracy 1/4 BPS black holes in four dimensional ${\cal N}=4$ supergravity. In this case, the object of interest is\footnote{ Note that the black hole degeneracy $d(\cal Q)$ has a factor of $(-1)^{l+1}$ relative to the definition (\ref{eq:sf1}). This factor arises from a careful treatment of the helicity quantum numbers when we go from five down to four dimensions \cite{Shih:2005he,ChengVerlinde2007}. }
\begin{equation}\label{eq:d1}
 d({\cal Q}) = (-1)^{l+1}\int_{\mathcal{C}}d\tau d\rho dz \,e^{-2\pi i \Qc\cdot Y}\frac{1}{\chi_{10}( \Omega)}~.
\end{equation}
where  $\chi_{10}$ is the Igusa cusp form. In the following we will summarise the procedure done in \cite{JatkarSen2006,David:2006yn,Sen:2007qy} to extract the asymptotic growth; this will serve as a guiding principle for the later cases. In the next example we will derive more general expressions and capture more broadly the data that governs the logarithmic corrections from the statistical point of view.

To start, it is useful to highlight some basic properties of $\chi_{10}$. As an {\it additive lift} we can write it as
  \be\label{eq:f10}
  \chi_{10}(\Omega) = \sum_{m=1}^\infty (\phi_{10,1}| V_m)(\tau,z)p^m~,
  \ee
  where
  \be
  \phi_{10,1}=\eta^{18}(\tau) \theta_1^2(\tau,z)~,
  \ee
and $V_m$ is  the Hecke operator in \eqref{eq:hecke}. For now it is not important the details behind $V_m$; in what follows, the important observation is that $\phi_{10,1}$ is the seed. The miracle of $\chi_{10}$ is that it can also be written as an {\it exponential lift}, which reads
  \bea\label{eq:exp10}
  \chi_{10}(\Omega) &=&\textrm{Exp-Lift}(2\phi_{0,1}) \cr
  &=& qyp\prod_{(r,s,t)>0}(1-q^sy^t p^r)^{2C_0(4rs-t^2)}~.
  \eea
  Here $C_0$ are the Fourier coefficients of   $\phi_{0,1}$ given in \eqref{eq:jfex}. What is physically interesting of this example is its intimate relation to the elliptic genera of $K3$. More concretely
    \bea
  \phi_{0,1}= {1\over 2}\chi(\tau,z;K3)~,
  \eea
  and hence another way to write the Igusa form is as
    \be\label{eq:10-TB}
  {1\over \chi_{10}(\Omega)}={\hat Z(\Omega)\over \phi_{10,1}(\tau,z)}~,
  \ee
  where
  \bea
  \hat Z(\Omega)&=& \sum_{m=-1}^{\infty} \chi(\tau,z; {\rm Sym}^{m+1}(K3))p^m\cr
  &=&  p^{-1}\prod_{r>0,s\geq 0,t}(1-q^sy^t p^r)^{-2C_0(4rs-t^2)}~.
  \eea
  Equation \eqref{eq:10-TB} has an interesting physics interpretation. As whole, \eqref{eq:10-TB} counts four dimensional 1/4-BPS dyons.  The factor of  $\hat Z(\Omega)$, while it is not a SMF, it is the counting formula for the Strominger-Vafa 5D black hole. The factor of $ \phi_{10,1}(\tau,z)$ arises from placing the 5D black hole on Taub-Nut: it counts bound states of the Kaluza-Klein monopole and the center of mass motion of the black hole. This is know as the 4D-5D lift \cite{Gaiotto:2005gf,Shih:2005uc,David:2006yn}. It is both remarkable and powerful that $\chi_{10}$ has the capacity to capture the degeneracy of BPS 4D and 5D configurations.

 We now discuss the asymptotic behavior of the Fourier coefficients \eqref{eq:d1}. As we outlined above, we will first perform a residue integral around the most dominant pole.  The positions of the zeros of $\chi_{10}$ are given by $H_1(1)$, and from \eqref{eq:dompole} we deduce that the most dominant pole in the regime \eqref{eq:largeQ} is given by
   \be\label{eq:10-pole}
  \tau\rho-z^2  +z  =0~.
  \ee
  As in \cite{Sen:2007qy}, it is convenient to map this pole to  $z=0$ for the simple reason that the residue there takes the form given in \eqref{liftresidue}. The $Sp(4,\Z)$ element that maps does the trick is  
  \be\label{Ashoketrafo}
\gamma =\twobytwo{A}{B}{C}{D}=\left(
\begin{array}{cccc}
 0 & 1 & -1 & 0 \\
 0 & 1 & 0 & 0 \\
 1 & 0 & 0 & 0 \\
 -1 & 0 & 0 & 1 \\
\end{array}
\right)~,
\ee
which acts as 
\be
\hat \Omega:=\gamma(\Omega)= (A\Omega+B)(C\Omega+D)^{-1}~,
\ee
and the individual components transform as 
\be\label{eq:hatvar}
\tau = \frac{1}{2\hat z-\hat \rho-\hat \tau}~,\qquad
\rho = \frac{\hat z^2-\hat \rho\hat \tau}{2\hat z-\hat \rho-\hat \tau}~,\qquad
 z = \frac{\hat z-\hat\rho}{2\hat z-\hat \rho-\hat \tau}~.
\ee
Under such a transformation \eqref{eq:10-pole} goes to $\hat z=0$ and the integrand \eqref{eq:d1} will change as
\bea
d({\cal Q})&=&(-1)^{l+1}\int_{\mathcal{C}}d\tau d\rho dz \,e^{-2\pi i \Qc \cdot Y}\frac{1}{\chi_{10}(\Omega)}\cr
&=& (-1)^{l+1}\int_{\cal C} d\hat\tau d\hat\rho d\hat z\, e^{-2\pi i \Qc\cdot Y} \det(C \Omega +D)^{3+k}{1\over \chi_{10}(\hat\Omega)}~.
\eea
We have used \eqref{eq:tz}, and for $\chi_{10}$ we have $k=10$. The Jacobian of the transformation \eqref{Ashoketrafo} is
\be
\det(C \Omega +D)^3=(2\hat z-\hat \rho -\hat \tau)^{-3}~.
\ee
Following \eqref{liftresidue}, near $\hat z=0$ we have
\be\label{eq:res10}
\frac{1}{\chi_{10}(\Omega)} = {1\over (2\pi i\hat z)^2}\eta(\hat\tau)^{-24A} \eta(\hat\rho)^{-24A}+\cdots~,
\ee
where 
\be
A= {1\over 24}\sum_{l} c(0,l)= {k+2\over 12}~.
\ee 
The first equality expresses $A$ in terms of $\phi_{0,1}$ (which is the seed in the exponential lift \eqref{eq:exp10}); the second equality highlights the fact that the weight of the SMF fixes the weight of the residue, that is the power of the $\eta$ functions. Note that here $m_{1,1}=2$ which follows from \eqref{eq:m11} and \eqref{eq:jfex}. 

Performing a contour integral around a contour $\cal C$ that encloses \eqref{eq:10-pole} gives
\be\label{eq:firstint}
d({\cal Q}) \approx {(-1)^{l}\over 4\pi^2}\int d\hat \tau d\hat \rho \, e^{-{2\pi i\over \hat \rho + \hat \tau}(m\hat \tau \hat \rho -n+l\hat \rho)}\, \eta(\hat\tau)^{-24A} \eta(\hat\rho)^{-24A} g_{\rm res}(\hat \tau,\hat \rho)~,
\ee
with
\bea\label{eq:gr}
g_{\rm res}(\hat \tau,\hat \rho)= {2\pi i}\le(e^{2\pi i\Qc\cdot Y} {d\over d\hat z} \le(e^{-2\pi i\Qc\cdot Y}\det(C \Omega +D)^{3+k} \ri)\ri)_{\hat z=0}~.
\eea
The derivative comes from the fact that we have a quadratic pole.
Here the symbol ``$\approx$'' reflects upon the fact that we are only considering the pole \eqref{eq:10-pole}; corrections to \eqref{eq:firstint} are exponential suppressed for large $\Qc$ and come from considering other integer values of $f$ in $H_1(1)$. These corrections are tractable, hence a generalization to an exact formula for $d(\Qc)$ is rather feasible.  

It is convenient to redefine variables in \eqref{eq:firstint}: we introduce complex variables $\tau_{1,2}$ which are defined as
\be
\hat \rho =: \tau_1 +i\tau_2 ~,\quad \hat \tau=:-\tau_1 +i \tau_2 ~,
\ee
and we have
\be\label{eq:dq1}
d({\cal Q}) \approx {1\over 4\pi^2}\int d \tau_1 d \tau_2 \, e^{{\pi \over \tau_2}(m (\tau_1^2+\tau_2^2) +n-l\tau_1)}\, \eta( -\tau_1 +i\tau_2)^{-24A} \eta( \tau_1 +i\tau_2)^{-24A} g_{\rm res}(\tau_1,\tau_2)~,
\ee
and \eqref{eq:gr} becomes
\be
g_{\rm res}(\tau_1,\tau_2)= -4\pi i (-2i\tau_2 )^{-(k+4)}\le((3+k)+ {\pi\over \tau_2} (n -l \tau_1 +  m (\tau_1^2 + \tau_2^2))\ri) ~.
\ee
Equation \eqref{eq:dq1} gives a systematic way on how to compare the  statistical degeneracy to a dual holographic system. In particular to extract the leading contribution and its logarithmic correction, the next and final step is to estimate this integral by a saddle point approximation.   

As $\Qc^2\gg1$, the position of the saddle point is governed by the explicit exponential term in \eqref{eq:dq1}, and its location is given by
\be\label{eq:t1t2}
\tau_1^*={l\over 2m}~,\quad \tau_2^*={1\over 2m} \sqrt{2\Qc^2}~,
\ee
and the leading contribution to \eqref{eq:dq1}  becomes
\be\label{eq:sd}
d({\cal Q}) \approx e^{\pi \sqrt{2\Qc^2} }   \eta( -\tau_1^* +i\tau_2^*)^{-24A} \eta( \tau_1^* +i\tau_2^*)^{-24A} g_{\rm res}(\tau_1^*,\tau_2^*) \le(2{(\tau_2^*)^2\over  \sqrt{2\Qc^2}}\ri)~,
\ee
where the last term in parenthesis is the contribution of the measure  in \eqref{eq:dq1}. 
From \eqref{eq:t1t2} one can see that for $\Qc^2\gg1$ the most dominant term is the explicit exponential factor in \eqref{eq:dq1}, and it  justifies our initial assumption in \eqref{eq:f-ext}. The $\eta$-functions, while they can contribute with exponential contributions if its entries are small, give subleading corrections in this large charge limit. This class of corrections to $d(\Qc)$ are as well interesting (they are usually interpreted as higher derivative corrections), but not our present focus; see, e.g.,  \cite{Sen:2007qy,Banerjee:2008ky} and references within.  

Note that the leading exponential in \eqref{eq:sd} is in agreement with \eqref{eq:z4}: for $\chi_{10}$ we have $D=1$ and $t=1$. This is the universal correction that in the gravitational language would be the ``area law'' and in the CFT it mimics the Cardy growth of states.  However, we want to make a contrast among these regimes and how subleading corrections are sensitive to them. In the following we will record the leading and subleading logarithmic correction in physically different scaling regimes for which \eqref{eq:sd} holds.
\begin{description}
\item[I. $n\gg 1$, $m\sim O(1)$]: Without loss of generality, it is convenient to introduce a scale $\Lambda\gg1$ and take
\be
n\sim \Lambda^2~,\quad m\sim \Lambda^0~,\quad l\sim \Lambda~. 
\ee 
 In this regime we have
\be
\tau_1^* \sim\Lambda ~,\quad \tau_2^*\sim \Lambda ~,\quad \Qc^2\sim \Lambda^{2}~, 
\ee 
and the asymptotic growth behaves as   
\bea\label{eq:1-10}
\ln d(\Qc)&\approx& \pi \sqrt{2\Qc^2} -12\ln \tau_2^* + \cdots \cr
&\approx& \pi \sqrt{2\Qc^2} -{12}\ln \Lambda + \cdots~,
\eea
where we included the logarithmic correction and used that $k=10$. Note that the logarithm in this scaling regime is independent of the eta-functions: only $g_{\rm res}$ and the measure in \eqref{eq:sd} contribute. This is in complete agreement with the expected correction in the Cardy regime in \eqref{eq:c4}. From the standpoint the Fourier decomposition in \eqref{eq:f10} of $\chi_{10}$, this correction is rather predictable since it is the universal contribution that is controlled by modular properties of $\phi_{10,1}$. From the standpoint of the exponential lift in \eqref{eq:exp10}, this data is more intricate: $k$ is controlled by the low lying coefficients in $\phi_{0,1}$. 

\item[II. $n\sim m\gg 1$]:   Here we set
\be
n\sim \Lambda~,\quad m\sim \Lambda~,\quad l^2\sim \Lambda^2~,
\ee
which gives
\be
\tau_1^* \sim\Lambda^{0} ~,\quad \tau_2^*\sim \Lambda^{0} ~,\quad \Qc^2\sim \Lambda^{2} ~.
\ee 
Since the moduli $\tau_{1,2}$ do not scale, the only logarithmic correction arises from the explicit dependence of $\Qc$ in \eqref{eq:sd}, i.e. from $g_{\rm res}$ and the measure factor. The degeneracy is
\bea\label{eq:cr1}
\ln d(\Qc)&\approx& \pi \sqrt{2\Qc^2} +(1-1)\ln \sqrt{\Qc} + \cdots\cr
&\approx& \pi \sqrt{2\Qc^2} + \cdots~.
\eea
This reproduces the results in \cite{Sen:2007qy,BanerjeeGuptaMandalEtAl2011}. Note that the logarithmic correction will vanish every time we have a pole of order 2 and $\tau_{1,2}$ do not scale with $\Lambda$. In general, this will be  the easiest regime to capture since the moduli are of order one.
 
\item[III. $m\gg n\gg 1$]: If now instead we take the $m$ to be arbitrarily large,  we have
\be\label{eq:n5d}
n\sim \Lambda~,\quad m\sim \Lambda^2~,\quad l^2\sim \Lambda^3 ~.
\ee 
 In this regime we have
\be
\tau_1^* \sim\Lambda^{-1/2} ~,\quad \tau_2^*\sim \Lambda^{-1/2} ~,\quad \Qc^2\sim \Lambda^{3} ~.
\ee 
The degeneracy is
\bea\label{eq:cr2}
d(\Qc)&\approx& \pi \sqrt{2\Qc^2} +12 \ln \tau_2^* + \cdots\cr
&\approx& \pi \sqrt{2\Qc^2} -12\ln \Lambda^{1/2} + \cdots ~.
\eea
In this regime all factors in \eqref{eq:sd} have a non-trivial contribution to the logarithm. Even though $d(\Qc)$ here needs to be modified to account for the statistical entropy of 5D black holes, this expression reproduces the logarithmic correction of the BMPV solution obtained in \cite{Sen:2012cj}.\footnote{And in addition, in \cite{Sen:2012cj} the scaling differed slightly from \eqref{eq:n5d} by including $l^2\sim \Lambda^{3+\alpha}$. This new parameter $\alpha$ is sensitive to the 4D-5D lift and hence it affects the coefficient in front of the log.}  It is also a coincidence, that \eqref{eq:cr1} and \eqref{eq:cr2} give the same correction; in our following examples we will generalize this result and the differences among regimes will be explicit. 
\end{description}
The devil is in the details. Each of these scalings regimes has a universal leading contribution, which is identified with the area law contribution in gravity. The logarithmic corrections are also crucial for this identification: a two derivative theory of gravity makes a prediction on both the exponential piece (which measures the size of the black hole) and the logarithmic piece (which captures the perturbative fluctuations of the theory). As we mentioned before, for $\chi_{10}$ and CHL models, the agreement is a remarkable test of quantum gravity.

\subsubsection{Exponential lifts of weak Jacobi forms}\label{sec:CY}

$\chi_{10}$ is just one of a larger class of SMF that serves  our purpose, i.e. the purpose of building counting formulas with ``black hole'' features.  In this subsection we will identify such SMFs and quantify the behavior of their Fourier coefficients.  These examples involve exponential lifts of weak Jacobi forms whose modular group is $\Gamma_t^+$. 

Our starting point is to consider
\begin{equation}\label{eq:dcy}
 d({\cal Q}) =\int_{\mathcal{C}}d\tau d\rho dz \,e^{-2\pi i \Qc \cdot Y}\frac{1}{\Phi_{k}( \Omega)}~,
\end{equation}
where $\Phi_{k}$ is of the form \eqref{explift} and $\varphi\in J^{weak}_{0,t}$. The zeros of $\Phi_k$ are given by \eqref{eq:hdb}, and we are assuming they are non-trivial. The procedure to obtain $d(\Qc)$ follows very closely those steps for $\chi_{10}$: we first do a contour integral over the most dominant pole which brings the integral to the form similar to  \eqref{eq:firstint} and then a saddle point approximation as in \eqref{eq:sd}. 

As we argued around \eqref{eq:ddq}-\eqref{eq:dompole}, the most dominant pole will be that with maximal polarity $D$.  And the simplest case is when the dominance is given by $D=1$ and $b=1$, i.e. the Humbert surface is $H_1(1)$, and it will be the focus of the remainder of this section.  The most dominant pole is then described by the curve
\be
t(\rho\tau-z^2)+ z=0~,
\ee
As we did for $\chi_{10}$ it is useful to map this pole to $\hat z=0$, where we have a simple expression for the residue. A $\Gamma_t^+$ transformation that accomplishes this is
   \be\label{Gt}
\gamma =\twobytwo{A}{B}{C}{D}=\left(
\begin{array}{cccc}
0 &  \sqrt{t}  & -{1\over \sqrt{t}}&0 \\
 0& {1\over \sqrt{t}}  & 0 & 0 \\
 {1\over \sqrt{t}}&0 & 0 & 0 \\
 - \sqrt{t} &0&0 &{1\over \sqrt{t} } \\
\end{array}
\right)~,
\ee
which acts in the individual components as 
\be
\tau = \frac{1}{2t\hat z-t^2\hat \rho-\hat \tau}~,\qquad
\rho = \frac{\hat z^2-\hat \rho\hat \tau}{2t\hat z-t^2\hat \rho-\hat \tau}~,\qquad
 z = \frac{\hat z-t\hat\rho}{2t\hat z-t^2\hat \rho-\hat \tau}~.
\ee
The contour integral around this pole will generically give (up to numerical factors)
\be\label{eq:firstintcy}
d({\cal Q}) \approx \int d\hat \tau d\hat \rho \, e^{-2\pi i \Qc\cdot Y(\hat z=0)}\, \eta(\hat\tau)^{-24A} \eta(\hat\rho)^{-24A} g_{\rm res}(\hat \tau,\hat \rho)~,\ee
where we have used that $\Phi_k$ around $\hat z=0$ takes the general form \eqref{liftresidue}. The difference relative to \eqref{eq:firstint} is that now we have
\be
A= {1\over 24}\sum_{l} c(0,l)= {k+m_{1,1}\over 12}~,  \qquad m_{1,1}= \sum_{n>0} c(0,n)~,
\ee
 which generalizes the order of the pole;\footnote{We are assuming that $m_{1,1}>0$; otherwise we would have a zero instead of a pole. This is however easily fix by taking in \eqref{eq:dcy} $\Phi_k$ instead of $1/\Phi_k$. } recall that here $c(n,l)$ are the Fourier coefficients of $\varphi$ in the exponential lift. And the remaining piece in \eqref{eq:firstintcy} is now given by
 \bea\label{eq:grcy}
g_{\rm res}(\hat \tau,\hat \rho)= {2\pi i\over (m_{1,1}-1)!}\le(e^{2\pi i\Qc\cdot Y} {d^{m_{1,1}-1}\over d\hat z^{m_{1,1}-1}} \le(e^{-2\pi i\Qc\cdot Y}\det(C \Omega +D)^{3+k} \ri)\ri)_{\hat z=0}~.
\eea
The rest of the steps from here follow very closely as those in section \ref{sec:chi10}. Changing variables to $\hat \rho=t^{-1}(\tau_1+i\tau_2)$ and $\hat\tau=t(-\tau_1+i\tau_2)$ gives an integral similar to \eqref{eq:dq1},  and after making a saddle point approximation we obtain
\be\label{eq:sdcy}
d({\cal Q}) \approx e^{{\pi\over t} \sqrt{2\Qc^2} }   \eta( -\tau_1^* +i\tau_2^*)^{-24A} \eta( \tau_1^* +i\tau_2^*)^{-24A} g_{\rm res}(\tau_1^*,\tau_2^*) \le({(\tau_2^*)^2\over \pi \sqrt{2\Qc^2}}\ri)~,
\ee
 where 
 \be\label{eq:sacy}
 \tau_1^* = {l\over 2m}~,\quad \tau_2^*={1\over 2m} \sqrt{2\Qc^2}~.
 \ee
As expected again, the leading exponential contribution in this case agrees with \eqref{eq:z4}.  

With these expressions we can easily extract the logarithmic corrections to $\ln d(\Qc)$. The results for the three cases of interest are in Table \ref{tb:logcy}. The two pieces of data that can affect the logarithmic correction is either the weight of $\Phi_k$, or the order of the pole. What is interesting about this result is how this data is controlled by the seed in the exponential lift: being capable of capturing these logarithmic corrections in a gravitational setup probes non-trivially $\varphi$. As we mentioned in section \ref{sec:explift}, $\varphi$ would be the starting point to give a physical interpretation of $\Phi_k$ as a generating function for a family of CFTs.  
\begin{table}[h]
\centering
{\renewcommand{\arraystretch}{1.5}%
\begin{tabular}{|c|c|c|c|}
\hline
{\bf Scaling regime} &$\tau_{1,2}^*$  & $\Qc^2$ & $\ln \Lambda$ \\ \hline\hline
{\bf I.} $n\gg 1$& \multirow{2}{*}{$\Lambda$}&\multirow{2}{*}{$\Lambda^2$} & \multirow{2}{*}{$-{(k+2)}$}\\  
$n\sim \Lambda^2$,  $m\sim O(1)$, $l\sim \Lambda$ &  & &\\ \hline 
{\bf II.} $n\sim m\gg 1$& \multirow{2}{*}{$\Lambda^0$} &\multirow{2}{*}{$\Lambda^2$} & \multirow{2}{*}{$m_{1,1}-2$}\\  
$n\sim \Lambda$,  $m\sim \Lambda$, $l\sim \Lambda$ &  & &\\ \hline 
{\bf III.} $m\gg n\gg 1$& \multirow{2}{*}{$\Lambda^{-1/2}$}&\multirow{2}{*}{$\Lambda^3$}  & \multirow{2}{*}{$m_{1,1}-3-{k\over2}$}\\  
$n\sim \Lambda$,  $m\sim \Lambda^2$, $l\sim \Lambda^{3/2}$ &  & &\\ \hline 
\end{tabular}}
\caption{Summary of logarithmic corrections to $\ln d(\Qc)$ for SMF built out of weak Jacobi Forms, and  with maximal polarity $D=1$. Here $\Lambda\gg1$ which controls the scaling of $\Qc$. }
\label{tb:logcy}
\end{table}

It might seem like restricting our attention to SMFs that have $H_1(1)$ as the most dominant divisor is too restrictive. There are however  examples of such forms and in particular for which the results in Table \ref{tb:logcy} applies. These examples were first presented in \cite{Gritsenko:1999fk}, and the weak Jacobi forms used in the exponential lifts are related to Elliptic Genera of Calabi-Yau manifolds. 
The idea is to take exponential lifts of weak Jacobi forms whose most polar terms have polarity $-1$ rather than $-t^2$. Geometrically this means that some of the Hodge numbers conspire to cancel the leading polar terms.
To describe these examples it is useful to introduce the quantity
\be
\chi_p(M) = \sum_j (-1)^j h^{j,p}(M)~,
\ee
where $h^{j,p}$ are the Hodge numbers of CY$_d$. Note that we are abusing a bit notation: we hope it is clear when $\chi_p$ refers to a topological invariant versus an example of a SMF or elliptic genera as used in other sections.   
\begin{description}
\item[CY$_6$:] As a first example, which is not related to $\chi_{10}$, consider
\be\label{eq:cy6}
\Phi_1(\Omega)= \textrm{Exp-Lift}(\, \phi_{0,3})~.
\ee 
The paramodular group is  $\Gamma_3^+$, the weight is 1, and the first few coefficients of the Jacobi form are
\be
\phi_{0,3}=\phi_{0,{3\over 2}}^2= y+2+y^{-1}+q(\cdots)~,
\ee 
with $\phi_{0,3/2}$ as defined in \eqref{eq:exwj}. 

If  $\chi_0=\chi_1=0$, then  the relation between $\phi_{0,3}$ is the elliptic genus for a  CY$_6$ is
\be
\chi(\tau,z)_{{\rm CY}_6} = -\chi_2 \, \phi_{0,3}(\tau,z)\ ,
\ee
For this specific class of Calabi-Yau manifolds we will have that the  divisor is just $H_1(1)$.

\item[CY$_4$:] Another example of SMFs is given by
\be\label{eq:cy41}
\Phi_2(\Omega)= \textrm{Exp-Lift}( \phi_{0,2})~.
\ee 
The paramodular group is  $\Gamma_2^+$, the weight is 2, and  $\phi_{0,2}$ is defined in \eqref{eq:exwj1}. 
Its relation to the Elliptic genus for a  CY$_4$  is
\be
\chi(\tau,z)_{{\rm CY}_4} = -\chi_1 \, \phi_{0,2}(\tau,z)\ ,
\ee
where  CY$_4$ has $\chi_0=0$. And as expected the only divisor in this case is $H_1(1)$.
\item[CY$_3$:] In this case we have
\be\label{eq:cy31}
\Phi_{(3)}(\Omega)= \textrm{Exp-Lift}(\phi_{0,{3\over 2}}(\tau,2z)) \quad .
\ee 
The paramodular group is  $\Gamma_6^+$, but it is important to note that the weight is \emph{zero} (the subscript `3' here refers to CY$_3$). What is interesting of this example is that the divisor is $H_1(1)-H_1(5)$: we have both a pole and a zero. In a sense, $\Phi_{(3)}$ is for $\Gamma_6^+$ what the $J$-function represents for $SL(2,\Z)$. Having a zero and a pole does not affect in an obvious manner our derivations, but it might be interesting to explore if such feature has any physical repercussions. 

The relation between $\phi_{0,{3\over 2}}$ to the elliptic genus of CY$_3$ is
\be
\chi(\tau,z)_{{\rm CY}_3} = \frac12 e(M) \phi_{0,3/2}(\tau,z)\ ,
\ee
where
\be
e(M)=\sum_{p=0}^3(-1)^p\chi_p(M)= 2(h^{1,1}-h^{2,1})~.
\ee
Note that depending on the sign of $e(M)$ we would either want to consider in \eqref{eq:dcy} $\Phi_{(3)}$ itself or its reciprocal. Assuming that $e(M)>0$, we have
\be
k={1\over 2} c(0,0)=0~,\quad m_{1,1}= \frac12 e(M)~, 
\ee
which is the data that governs the logarithmic corrections. While $\Phi_{(3)}$ is not the counting formula for ${\cal N}=2$ BPS black holes in 4D, it is interesting to compare the coefficients of the log corrections;  the results in \cite{Sen2012b}  predict that logarithmic correction to the black hole entropy is $(2-e(M)/24)\ln\Lambda^2$ which does not match any of the regimes listed in Table \ref{tb:logcy}.  
\end{description}

Before addressing other SMFs, let us discuss briefly what are potential differences and obstacles if we have a form where the most dominant Humbert surface is not $H_1(1)$. 
So far, there are two important technical features in our derivations: identifying the most dominant pole, and the explicit expressions of the residue around that pole. The first feature is straightforward and transparent, which is outlined in \eqref{eq:f-ext}-\eqref{eq:dompole}. However, there is an important issue that we have not addressed so far.  Basically we need to discuss our choice of contour that encloses this pole and if this imposes significant restrictions on the saddle point.  

In a nutshell, our contour is restricted by the convergence of the expansion of the SMF. This is important since we are taking reciprocals of cusp forms which contain poles and we have to decide on which side of the pole we stand. Fortunately, the potential restrictions for $H_1(1)$ are rather simple: we are expanding $1/\Phi(\Omega)$ around $z=0$, and to guarantee convergence we choose  
\be\label{eq:yc}
|y|<1 \qquad \Rightarrow \qquad {\rm Im} z>0~.
\ee 
Our contour $\cal C$ has to lie within this domain, and therefore any further manipulation of the variables has to be compatible with this restriction. In particular, our saddle point \eqref{eq:ex1} needs to be compatible with \eqref{eq:yc}, which requires 
\be\label{eq:zl1}
{\rm Im} z_{\rm max} = {l\over 2 t f |\lambda|}  >0 ~.
\ee
Therefore, our derivations so far only apply if the $U(1)$ quantum number is positive. But this is rather mild condition that does not tamper with the main portion of our results in Table \ref{tb:logcy}. More generally, the convergence condition \eqref{eq:yc} depends on the Humbert surface in play. As we will see in our next example, the specification of the contour dramatically tampers with the growth in $d(\Qc)$.

Our second obstacle is the residue at a given pole. In certain cases, such as $H_1(1)$ and $\varphi\in J^{weak}$, we can write simple expressions such as \eqref{liftresidue} which allow us to derive universal results for $d(\Qc)$ that are applicable in a wide regime of charges. For general Humbert surfaces the task is more difficult. In the next subsection we will show how we can overcome some of these difficulties for Humbert surfaces of the type $H_D(0)$. 

\subsubsection{$\chi_{35}$}
Let us now return to classical SMF.
The Igusa modular form $\chi_{35}$ is the first SMF of odd weight with respect to $\Gamma_1=Sp(4,\Z)$. The most common definition of $\chi_{35}$ is given in terms of a theta series which can be found in e.g. \cite{MR2385372}.
For our purposes it is better to write it as an exponential lift  \eqref{explift}
as in \cite{MR1616929}. Explicitly we have  
\bea\label{eq:35exp}
\chi_{35}&=&\textrm{Exp-Lift}(\varphi_{0,1}^{(2)})(\Omega)\cr
&=& q^3 y p^2\prod_{(n,l,m)>0}(1-q^n y^l p^m)^{f_1^{(2)}(4nm-l^2)}~.
\eea
Here the seed in the lift is the nearly holomorphic
Jacobi form
\be
\varphi_{0,1}^{(2)}=(T_2 -2)\phi_{0,1}~,
\ee
with $T_2$ the Hecke operator \eqref{eq:hecke}. That is, the function $c^{(2)}(mn,l)=f^{(2)}_1(4mn-l^2)$ is given by the Fourier coefficients of  $\varphi_{0,1}^{(2)}$. We can evaluate \eqref{eq:hecke} acting on
$\phi_{0,1}$ explicitly to obtain 
\be
f^{(2)}_1(N)=8f_1(4N)+2(\left(\frac{-N}{2}\right)-1)f_1(N)+\delta^{(4)}_{0,N} f_1(N/4)\ ,
\ee
where $\delta^{(k)}$ is the periodic Kronecker delta, $\left(\frac{-N}{2}\right)$ is the Kronecker symbol and $f_1(N)$ are the Fourier coefficients of $\phi_{0,1}$. 

 The exponential lift \eqref{eq:35exp} is rather interesting. Although the Hecke operators $T_p$ map Jacobi forms to Jacobi forms of the same weight and index, their action on weak Jacobi forms is not as nice. In particular, they do not map weak Jacobi forms to weak Jacobi forms. For example, the function $\varphi_{0,1}^{(2)}$ has coefficients with $4n-l^2<-1$, and therefore is \emph{not} holomorphic. However, it is a \emph{nearly} holomorphic Jacobi form. Its first few Fourier coefficients are given by
\be
\varphi_{0,1}^{(2)}(\tau,z)=q^{-1} + y^{-2} + 70 + y^2 +q\left( 70 y^{-2} + 32384 y^{-1}+131976 + 32384 y + 70 y^2 \right)+\cdots~.
\ee
The generator $\chi_{35}$ can thus be written as the exponential lift of a nearly holomorphic Jacobi form. Moreover,  for $y=1$, we have
\bea\label{phi2Witten}
\varphi_{0,1}^{(2)}(\tau,0)&=&  q^{-1} + 72 +196884 q+ 21493760 q^2 +\cdots \cr
&=&72+ J(q)~,
\eea
with $J(q)$ the $J$-function. This gives an elegant tie of $\chi_{35}$ to near-extremal CFTs as defined \cite{Benjamin:2016aww} which would be interesting to study further. 
The first few Fourier coefficients of $\chi_{35}$ are
\bea
\chi_{35}&=&q^2y p^2(q-p)\Bigg[ 1-y^{-2}+q\left(y^{-4}+69y^{-2}-69-y^2\right) \\
&+&p\Big(y^{-4}+69y^{-2}-69-y^2+q( -y^{-6}+32384y^{-3}+129421y^{-2}-129421-32384y+y^4)\Big)+\cdots\Bigg]~, \notag
\eea
and the first few terms of its Fourier-Jacobi decomposition are
\bea
\chi_{35}&=& p^2\left( \frac{\xFive}{5159780352}\left(\xOne^3-\xTwo^2\right)^3 \right)  \cr
&&\cr
&+& p^3\left( \frac{\xFive\left(\xOne^3-\xTwo^2\right)^2}{644972544}\left(18\xOne^2\xTwo \xThree - 11\xOne^3\xFour - 7 \xTwo^2 \xFour \right) \right)   
+\cdots~, 
\eea
%
where $E_{4,6}$ are the Eisenstein series with $\xOne^3-\xTwo^2=1728 \Delta$ and the other Jacobi forms are defined in \rref{eq:jfex}.


In relation to our goal, the question is: Could $\chi_{35}$ count the entropy of a black hole? Or more broadly, could it have a gravitational (or stringy) interpretation? To answer that,  as for the other examples, let us consider the asymptotic growth of
\begin{equation}
 d({\cal Q}) = \int_{\mathcal{C}}d\tau d\rho dz \,e^{-2\pi i \Qc\cdot Y}\frac{1}{\chi_{35}( \Omega)}~.
\end{equation}
Recall that we take the reciprocal so that we can have exponential growth in the Cardy regime. To estimate the growth of $d(\Qc)$ we need to analyze its divisor, which are given by the surfaces $H_1(1)$ and $H_4(0)$. For $H_1(1)$, the pole is in the orbit of $z=0$ and around there we have\footnote{We note that there is a typo in \cite{Igusa35} for the residue of $\chi_{35}$. The steps to derive this residue follow closely from those in \cite{BORCHERDS199030}. }
\be\label{eq:z35}
\chi_{35} = 4\pi i z\, \eta(\tau)^{72} \eta(\rho)^{72} (J(p)-J(q)) + \cdots~.
\ee
While this has an elegant structure, it is not the dominant pole as $\Qc^2\gg1$. Nonetheless, as a side remark note that the residue of $1/\chi_{35}$ is governed by the $J$-function and powers of the
$\eta$-function; this will give a drastically different behavior for $d(\Qc)$ relative to $1/\chi_{10}$, as we will show below. 

In the asymptotic regime the focus has to be on $H_4(0)$ (the Humbert surface with maximal discriminant) for which the relevant poles are the images of 
\be\label{eq:pqpole}
p=q~.
\ee
However, relative to a pole at $y=1$ and the analysis around \eqref{eq:yc}, this pole is more subtle: our contour is restricted by the convergence of the expansion which will affect dramatically $d(\Qc)$. Given \eqref{eq:35exp}, we want to expand $1/\chi_{35}$ in the regime\footnote{We could as well be on the other side of the pole \eqref{eq:pqpole} by choosing instead $\le| {q p^{-1}}\ri| <1$. Physically we are making a choice if either $p$ or $q$ capture the polar contribution of the CFT. Regardless of this choice our results are unchanged. } 
\be\label{eq:ineq}
\le| {p q^{-1}}\ri| <1\qquad \Rightarrow \qquad {\rm Im}\rho> {\rm Im}\tau~.
\ee
The  inequality clearly breaks the exchange symmetry  $\rho\leftrightarrow\tau$. Throughout our approximations to estimate $d(\Qc)$ we have to respect \eqref{eq:ineq}. In particular, the saddle point \eqref{eq:ex1} has to be compatible with this inequality, and this leads to
 \be\label{eq:boundnm}
 n>m~.
 \ee
This sharpens our second regime from $n\sim m$ to a strict inequality. It won't be impossible to access our third regime, when $m\gg n\gg 1$; as we will see below it will just require a more detailed inspection of the contour and the residue. 

For $n>m$,  we can proceed as we did before with our approximations and test their validity. An interesting feature, absent in other examples, is that in the case of $\chi_{35}$ special care is needed because the residue will have additional poles at finite values of $(\rho,\tau,z)$.   To proceed, lets evaluate the contribution of the surface $H_4(0)$. Using \eqref{eq:dompole} we need to integrate around
\be\label{eq:p35}
(\tau\rho-z^2) + 1 =0~,
\ee
where we used that for $\chi_{35}$ we have $t=1$, $D=4$, $b=0$ and $e=1$. Solving for $\rho$ and performing the contour integral gives
\be
d(\Qc)\approx 2\pi i \int_{\mathcal{C}} d\tau dz \exp\le(-2\pi i n \tau -2\pi i l z-{2\pi i m  z^2\over \tau }+{2\pi i m\over \tau}\ri) f_{\rm res}(\tau,z)  ~,
\ee
where $f_{\rm res}(\tau,z)$ is the residue of $1/\chi_{35}$ around \eqref{eq:p35}.  We choose a contour $\mathcal{C}$ such that one has $\text{Im}(\rho)>\text{Im}(\tau)$.
If we do a saddle point approximation, where we assume that the integral is dominated by the explicit exponential factor, we obtain
\be\label{eq:d35}
d(\Qc) \sim 2\pi i\, e^{4\pi\sqrt{{\cal Q}^2/2}}   f_{\rm res}(\tau_\star,z_\star) \le(\tau_\star^2\over2m \ri)~,
\ee
with
\bea\label{eq:s35}
\tau_\star= i\sqrt{2m^2 \over \Qc^2}~,\quad z_\star= -{l\tau_\star\over2 m}~.
\eea
We can easily see that the constraint $\text{Im}(\rho)>\text{Im}(\tau)$ implies $n>m$.

One might be concerned about  the behavior of the residue at \eqref{eq:p35} and hence the validity of \eqref{eq:d35}. It is a difficult problem to extract an exact formula for the residue, so the best we can do at this stage is to proceed as follows. First we expand $\chi_{35}$ in powers of $\hat{\rho}=\rho-(z^2-1)/\tau$, that is,
\begin{equation}
\chi_{35}(\Omega)=h(\frac{z^2-1}{\tau},\tau,z)\hat{\rho}+\mathcal{O}(\hat{\rho}^2)~.
\end{equation}
The residue is then simply 
\begin{equation}
f_{\text{res}}(\tau,z)=\frac{1}{h\left(\frac{z^2-1}{\tau},\tau,z\right)}~.
\end{equation}
On the other hand we can expand $\chi_{35}$ first in powers of $z$ and then in powers of $\hat{\rho}$, that is,
\begin{equation}
\chi_{35}(\Omega)=4\pi i\Delta(-1/\tau)^3\Delta(\tau)^3J'(-1/\tau)z \hat{\rho}+\mathcal{O}(z^2,\hat{\rho}^2)~, \quad \Delta(\tau)= \eta(\tau)^{24}~,
\end{equation}
where we used the fact that $\hat{\rho}=\rho+1/\tau+\mathcal{O}(z^2)$ and \eqref{eq:z35}. This implies that we must have
\begin{equation}
h\left(\frac{z^2-1}{\tau},\tau,z\right)=4\pi i\Delta(-1/\tau)^3\Delta(\tau)^3J'(-1/\tau)z+\mathcal{O}(z^2)~,\qquad z\ll 1~.
\end{equation}
where $J'(\tau)$ is the derivative of the $J$-function. Moreover, using the fact that 
\begin{equation}
J'(-1/\tau)=\tau^2J'(\tau)~,
\end{equation}
we have $J'(i)=0$, and therefore $f_{\text{res}}(\tau,z)$ has a pole at $(z,\tau)=(0,i)$. At this point we also have $\rho=i$ and thus the pole $(z,\tau)=(0,i)$ lies precisely at the boundary $\text{Im}(\rho)=\text{Im}(\tau)$.

 Since the contour $\mathcal{C}$ is chosen to lie inside the region $\text{Im}(\rho)>\text{Im}(\tau)$, for $n>m$ we really don't need to know the exact expression for $f_{\rm res}$ if we only want to estimate the leading growth and its logarithmic correction. In the intermediate regime where $n\gtrsim m$ (close to the inequality \eqref{eq:boundnm}) the moduli near the saddle point \eqref{eq:s35} does not scale as $m\sim \Lambda$. And in the regime $n\gg m$, the robustness and universality of the Cardy regime guarantees that the residue cannot affect the position of the saddle, which implies that $f_{\rm res}$ should not diverge as $\tau_\star \to i 0^+$. Therefore, for $n>m$ the growth will be exponential as in \eqref{eq:d35}; and the logarithmic corrections will be dominated by the weight ($k=35$) as $n\gg m$, and by the order of the pole ($m_{4,0}=1$) for $n\gtrsim m$. 

However when $m>n$ one has that  $\text{Im}(\rho)<\text{Im}(\tau)$ at the saddle point (\ref{eq:s35}). Therefore one has to deform the initial contour $\mathcal{C}$ to pass through the new saddle point and as a consequence it will have to cross the boundary $\text{Im}(\rho)=\text{Im}(\tau)$. Since the pole of $f_{\text{res}}(\tau,z)$ at $(z,\tau)=(0,i)$ lies precisely at this boundary, when deforming the contour we will pick the contribution of this pole, and thus
\begin{equation}
d(\Qc) \sim 2\pi i\, e^{4\pi\sqrt{{\cal Q}^2/2}}   f_{\rm res}(\tau_\star,z_\star)\le(\tau_\star^2\over2m \ri)-2\pi^2\frac{e^{2\pi(m+n)}}{\Delta(i)^6J''(i)}~,\qquad m>n~.
\end{equation}
For $m\sim n$, with $m>n$,  $\tau_\star$ and $z_\star$ are $\mathcal{O}(1)$ and so $f_{\rm res}$ cannot become large. Moreover, since we have the strict inequality $m+n>2\sqrt{{\cal Q}^2/2}$, we can approximate
\begin{equation}
d(\Qc) \sim e^{2\pi(m+n)}~,\qquad m>n\gg 1,
\end{equation} 
which shows that for $m>n$ the degeneracy has Hagedorn growth instead of Cardy growth, characteristic of the regime $n>m$.

\subsubsection{$\chi_{12}$}\label{sec:x12}
As a last example we now turn to $\chi_{12}$: this is a SMF of weight 12 under $Sp(4,\Z)$. It is a cusp form defined by
\be
\chi_{12}= \frac{1}{\mathcal{N}}\left(E_{12}^{(2)} - (E_6 ^{(2)})^2\right)~,
\ee
where $E_{12,6}^{(2)}$ are the Eisenstein series of genus two defined in \cite{MR2385372} and $\mathcal{N}$ is a normalization such that the coefficient of $qpy$ is set to one. The first few coefficients are
\be
\chi_{12}= p \bigg( q(y^{-1}+12 + y)+ q^2 (10 y^{-2} - 88 y^{-1} -132 - 88 y + 10 y^2) \bigg)+ \cdots~,
\ee
and the few terms in the Fourier-Jacobi expansion of $\chi_{12}$.
\bea \label{fourierjacobichi12}
\chi_{12}&=&p\frac{\xOne^3-\xTwo^2}{1728}\xFour+  p^2\left( \frac{\xOne^3-\xTwo^2}{864}\left( 6 \xOne \xThree^2 - \xFour^2\right) \right)\cr&& -p^3\left( \frac{\xOne^3-\xTwo^2}{6912}\left( 63 \xOne \xThree^2 \xFour - 60 \xTwo \xThree^3 - 7 \xFour^3\right) \right)   
+\cdots~.
\eea

To our knowledge, $\chi_{12}$ cannot be written as a exponential lift of the form \eqref{explift}. However, we can still use it to build counting formulas with `black hole' features. In \cite{Igusa} it is shown that $\chi_{10}$ and $\chi_{12}$ do not share any zeroes.\footnote{We thank Miranda Cheng and Gerard van der Geer for discussions on this point.} This in particular implies that the Fourier coefficients of a combination such as
\be
{\Phi}(\Omega)={\chi_{12}\over \chi_{10}}~,
\ee
will have the desired features. The strategy taken in section \ref{sec:chi10} still applies with only minor modifications: we will have 
\be
d({\cal Q}) \approx {(-1)^{l}\over 4\pi^2}\int d\hat \tau d\hat \rho \, e^{-{2\pi i\over \hat \rho + \hat \tau}(n\hat \tau \hat \rho -m+l\hat \rho)}\, g_{\rm res}(\hat \tau,\hat \rho)~,
\ee
where we integrated over the pole $\hat z=0$, and the hatted variables are given in \eqref{eq:hatvar}. In contrast to \eqref{eq:firstint}, note that we don't have a contribution from the residue since near $\hat z=0$ the behavior is \cite{Igusa} 
\be
\chi_{12}=\eta(\hat\tau)^{24A} \eta(\hat\rho)^{24A} +\cdots~,
\ee
and hence the contribution from the residue in \eqref{eq:res10} cancels against $\chi_{12}$. The results for the asymptotic behavior are also very simple: following the results in Table \ref{tb:logcy} we have $k=-2$ and $m_{1,1}=2$. Therefore, for all three scaling regimes we find
\be
\ln d(\Qc)\sim {\pi} \sqrt{2\Qc^2} + 0\times \ln \Qc ~, \qquad \Qc\gg1~.
\ee

One important assumption we are making in this example is that the contour of integration is further restricted by the addition of $\chi_{12}$ relative to the one used for $\chi_{10}$. We have not found evidence of such restriction, but we do not have a rigorous proof.

\section{Physical interpretation of SMFs}
 
 In this section we will discuss physical interpretation of SMFs from a CFT$_2$ perspective and a gravitational perspective. On the CFT side the emphasis will be on how and when can  we interpret our examples in terms elliptic genera of SCFTs. On the gravitational side we will suggest how we could read off more detailed information about the gravitational theory besides its black hole features.
 
\subsection{CFT origin of a SMF}\label{sec:cft}
We would now like to give a physical interpretation
to at least some of the SMFs that we
have been discussing. For those that can be cast as an exponential lift, the interpretation is simple: as illustrated by  \eqref{eq:symprod}, we can easily interpret it as the generating function of symmetric products (up to the contribution of possible prefactors), and the only challenge is to interpret the Jacobi form that enters in the exponential lift.   In this section we want to deviate from this class of examples. Without resorting to a product expansion of a SMF, we  want  to discuss 
if it is possible to interpret them as generating functions
of generalized partition functions such as the elliptic
genus of families of SCFTs. 

As discussed in the section \ref{sec:smf}, the key observation is
that the coefficients of the Fourier-Jacobi expansions
of a SMF of weight $k$,
that is its expansion in $p$, are Jacobi forms of weight
$k$ and index $m$. It is thus natural to try to interpret those
forms as for instance the elliptic genera of a family
of CFTs. An immediate problem however is that the
Jacobi forms have weight $k$, whereas we want
forms of weight 0. To address this we can try
to pull out an overall prefactor of weight $k$.
More precisely,
we define a Siegel Modular Form ${\Phi}$ (possibly of negative weight) to be of \textit{Elliptic Genera type} if it can be written as
\be
{\Phi}(\Omega)= p^{\ell}\mathcal{M}(q,y) \sum_{m=0}^\infty p^m \varphi_{0,m}(q,y)~,
\ee
for some holomorphic weak Jacobi forms $\varphi_{0,m}(q,y)$ that have zero weight and index $m$ and for some prefactor $p^\ell \mathcal{M}(q,y)$ with $\ell\in\mathbb{Z}$. Note that $\mathcal{M}(q,y)$ has weight $k$ under
$SL(2,\Z)$ transformations. If a SMF is of Elliptic Genera type, it can naturally be interpreted as coming from a family of supersymmetric 2d CFTs theories. The SMF is then built by taking the generating function of the elliptic genus and multiplying it with some prefactor $p^\ell\mathcal{M}(q,y)$.
The prefactor can sometimes be given a physical interpretation as for $\chi_{10}$. We now discuss this property in a class of examples.
First, we consider holomorphic SMF, which are in the ring given in (\ref{eq:ring}).

The Fourier-Jacobi expansion of $\chi_{10}$ is given by
\be\label{ch10FJ}
\chi_{10}= 
\frac{p}{1728} \left( \xOne^3 - \xTwo^2\right)\xThree\left(1 -  2p \xFour+\cdots \right)
\ ,
\ee
as can be seen from the exponential lift expression.
The prefactor $p^\ell\mathcal{M}(q,y)$ is thus $p \phi_{10,1}(\tau,z)$ (see \eqref{eq:jfex}),
and we can indeed get a family of weak Jacobi forms of weight 0.
Note that this is simply a consequence of the fact that
$\chi_{10}$ can be written as an exponential lift,
and the prefactor is the first factor in (\ref{explift2}).
A slightly different example is $\chi_{12}$, which
to our knowledge can not be written as an exponential
lift.
Unlike $\chi_{10}$,  the first term of its Jacobi-Fourier is not a prefactor of all other terms. However, note that 
\be\label{ch12FJ}
\chi_{12}= \frac{\xOne^3 - \xTwo^2}{1728}\left(p \xFour + 2 p^2 (6\xOne \xThree^2-\xFour^2)+\cdots \right)~, 
\ee
and since $\chi_{12}$ is a cusp form, in general, we can consider  
\be \label{chi12overD}
Z:=\frac{\chi_{12}}{\Delta(\tau)}\ , \qquad \Delta(\tau)=  \eta(\tau)^{24}=~\frac{\xOne^3 - \xTwo^2}{1728},
\ee 
where $Z$ is holomorphic and hence each term in the $p$ expansion will be a weak Jacobi form of weight 0 and increasing index. These Jacobi forms are constrained by the fact that $\chi_{12}$ was holomorphic and that it had an expansion in terms of Jacobi forms with no polar terms, which therefore only have polynomial growth. 
The exponential growth thus comes from $\Delta(q)$ in the denominator.
Using this fact, one can show that the coefficients of the weak Jacobi forms of \rref{chi12overD} obey the property
\be
c(4mn -l^2)=0, \qquad 4mn-l^2<-4m\ .
\ee
This means they don't have the most polar terms as would be allowed for a generic weak Jacobi form of that index. As a consequence, the asymptotic growth of the coefficients is given by \eqref{eq:c4} with $n_0=-1$ and $l_0=0$, namely
\be
c(4mn-l^2)\sim e^{4\pi\sqrt{\left(n-\frac{l^2}{4m}\right) }}
\ee
which is much slower than the usually Cardy type growth. 
Giving these weak Jacobi forms a CFT interpretation, it means for instance the vacuum does not contribute to the elliptic genus. It would be interesting to see what type of gravity interpretation can be given to such an object.

The issue with holomorphic SMF is thus that we will never
get proper Cardy growth that increases with the index $n$. To get
such exponential growth, we need to consider meromorphic SMF. The simplest
types of examples are reciprocals of a holomorphic SMF. 
For example, in the previous section we saw that $1/\chi_{10}$ was a SMF of elliptic genera type. More precisely,
we can use (\ref{ch10FJ}) to write 
\bea
\frac{1}{\chi_{10}}
&=&\frac{1}{ \frac{p}{1728} \left( \xOne^3 - \xTwo^2 \right)\xThree}\frac{1}{1-2p \xFour+\cdots} \notag \\
&=& \frac{1}{ \frac{p}{1728} \left( \xOne^3  - \xTwo^2 \right)\xThree} \left(1+ 2p  \xFour+\cdots\right) = \frac{\hat Z}{p\phi_{10,1}}\label{chi10exp}
\eea
Because we could pull out the factor of $(\xOne^3-\xTwo^2)\xThree$, the $p$ expansion in the parentheses of (\ref{chi10exp}) only contains positive powers of the generators and is therefore a holomorphic weak Jacobi form. In fact, for $\chi_{10}$ we know this is precisely the generating function of the symmetric orbifold of K3. The prefactor in this case has a physical intrepretation as counting degrees of freedom coming from the KK monopole as well as the center of mass modes.

Note that it was crucial here that we could pull out an overall prefactor
which left the remaining expansion starting as $1+O(p)$. This allowed us
to use the geometric series to invert the denominator. This procedure would
not work for instance for $1/\chi_{12}$: From (\ref{ch12FJ}) we see that the denominator contains a factor of $\xFour$, which, when expanding, would lead
to higher and higher poles in the Jacobi forms of the expansion.
In fact one may ask whether the reciprocal of any other holomorphic SMF gives a form of elliptic genera type. We have checked this explicitly for any element of the ring up to weight 20 
and none of those SMF have such a property. 

We can however consider examples of the form
\be
\frac{\chi_{12}}{\chi_{10}}= \frac{1}{p\, \phi_{-2,1}} \sum_m p^m \varphi_{0,m}\ .
\ee
For some weight zero weak Jacobi forms of increasing index. These will have an exponential growth of the form
\be
c(4mn-l^2)\sim e^{4\pi\sqrt{nm}} \,.
\ee
coming mostly from $1/\chi_{10}$ although $\chi_{12}$ will give potentially interesting subleading corrections. The weak Jacobi forms are not exactly
the symmetric orbifolds of $K3$. It would be very interesting to give them a physical interpretation. For example, it would be interesting to check whether they correspond to an orbifold by a different oligomorphic permutation group \cite{Belin:2014fna, Haehl:2014yla,Belin:2015hwa}.

Finally, note that $\chi_{35}$ is a somewhat special example. It is not of elliptic genera type in the sense we defined it here, but it can still be given a natural CFT interpretation. From the exponential lift we know that the family of CFTs is again a symmetric orbifold of a nearly holomorphic Jacobi seed form rather than a weak Jacobi form. If we are willing to work with such nearly holomorphic forms, we can then give an interpretation to both $\chi_{35}$ and its reciprocal.

\subsection{The gravitational dual of a SMF}\label{sec:gravint}
We have provided examples of SMFs that have an extended Cardy regime. This suggests that we could attribute this growth to a black hole within the natural ranges where a semi-classical gravitational description is valid. The question is, if we can identify the specific gravitational dual theory. In this section we suggest how to give a more refined bulk interpretation of our results, and in particular how  to interpret the residue formula. This is best understood for $\chi_{10}$. We will review previous work on the spectrum of chiral primaries on $AdS_3$ \cite{deBoer:1998us,Gaiotto:2006ns} and its interpretation as gas of BPS multiparticle states. From here we would like to suggest a similar interpretation for our other examples, and leave it as future work to test this interpretation.


For exponential lifts of weak Jacobi forms for which the dominant pole is in $H_1(1)$, we found the residue formula (\ref{liftresidue}) for the Fourier coefficients. 
Since this is an exact formula, we can study further corrections to the leading saddle point contribution, and try to extract information on the spectrum of the gravity dual. 
In particular, the pole at $\hat z=0$ and its images under modular transformations has the special feature that their residue factorize. Near $\hat z=0$ one finds
\begin{equation}
 \frac{1}{\Phi_k(\hat \Omega)}= {1\over \hat z^{m_{1,1}}} \eta(\hat \tau)^{-24A} \eta(t\hat\rho)^{-24A}+\cdots~, \qquad A={k+m_{1,1}\over 12}
 \end{equation}
 where $k$ is the weight of the modular form $\Phi_k$ and $m_{1,1}$ is the order of the pole. We can recast \eqref{eq:firstintcy} as 
 \begin{eqnarray}\label{eq:res}
d({\cal Q}) \approx \int d\tau_1 d\tau_2 \, e^{{\pi\over t \tau_2}(n-l\tau_1+m(\tau_1^2+\tau_2^2)) }\, \eta(t(-\tau_1+i\tau_2))^{-24A} \eta(\tau_1+i\tau_2)^{-24A} g_{\rm res} (\tau_1,\tau_2)
\end{eqnarray}
where we have neglected the contribution coming from the other poles, $g_{\rm res}$ is given by \eqref{eq:grcy}, and we have defined $\hat\rho=t^{-1}(\tau_1+i\tau_2)$ and $\hat \tau=t(-\tau_1+i\tau_2)$.
%
%
For simplicity we scale all charges uniformly, i.e. $m\sim n \sim \Lambda^2\gg1$, so that  $\tau_{1,2}\sim \mathcal{O}(\lambda^0)$. In this regime we can approximate 
\be
g_{\rm res} \sim \sqrt{\Qc^2 }^{m_{1,1}-1}\tau_2^{-1-k-m_{1,1}}~,
\ee
so that 
\begin{eqnarray}\label{res form}
d({\cal Q}) \sim  \int d\tau_1 d\tau_2 \, \tau_2^{-1-k-m_{1,1}} e^{{\pi\over t \tau_2}(n-l\tau_1+m(\tau_1^2+\tau_2^2)) }\, \eta(t(-\tau_1+i\tau_2))^{-24A} \eta(\tau_1+i\tau_2)^{-24A} ~,
\end{eqnarray}
where we ignored overall factors of $\Qc^2$. This formula contains more information than just the leading black hole entropy area formula. To illustrate how much is captured by it, let us discuss the case of $\chi_{10}$, for which $t=1$ and $A=1$. 

An interesting portion to interpret from this formula is the infinite products in the $\eta$-functions. For $\chi_{10}$ the infinite product arises from a trace over a gas of multiparticle BPS particles: in \cite{Gaiotto:2006ns} they show that this piece comes from contributions of the 5D supergravity multiplets which includes the graviton, vectors and hypermultiplets and also from (anti)-M2 branes wrapping holomorphic two cycles on the Calabi-Yau. 

Lets discuss in a bit more detail how the analysis of  \cite{Gaiotto:2006ns} breaks down for M-theory on $AdS_3\times S^2 \times M$ with $M$ a Calabi-Yau 6-fold. Here one has thermal $AdS_3$.  To relate to the black hole problem we have to perform a modular transformation, which takes the complex structure of the boundary torus $\tau$ to $-1/\tau$. For the extremal black hole that we are interested we have $\tau=i\phi^0$, where $1/\phi^0$ can be identified with the radius of the M-theory circle in the near horizon geometry (\ref{near horizon geometry}) of the black hole. The analysis of the supergravity modes is standard: the massless spectrum consists of the graviton multiplet, $n_V=h^{1,1}-1$ vectormultiplets and $n_H=2(h^{2,1}+1)$ hypermultiplets. The spectrum on $AdS_3\times S^2 \times M$ organizes into short representations of $SL(2,\mathbb{R})\times SU(1,1|2)$ that can be found in  \cite{Fujii:1998tc,deBoer:1998us}. The trace that one obtains after summing over the supergravity modes is 
\begin{equation}\label{eq:zsugra}
 Z_{sugra}=\prod_{n=1}^{\infty}(1-\zeta^n)^{-n \chi(M)}~,\,\zeta=e^{-\frac{2\pi}{\phi^0}}
\end{equation}
 where $\chi(M)=2(h^{1,1}-h^{2,1})$ is the Euler character of the Calabi-Yau manifold $M$.

The most interesting part comes from tracing over the BPS states due to (anti)-M2 branes wrapping holomorphic two cycles in $M$. This can be shown to equal the Gopakumar-Vafa BPS invariants partition function \cite{Gaiotto:2006ns}: that gives
\begin{equation}\label{eq:zm2}
 Z_{M2}=\prod_{n_a>0,k>0}\left(1-\zeta^{k}e^{-2\pi n_a \frac{t^a}{\phi^0}}\right)^{kd^0_{n_a}}\prod_{n_a>0,r>0}\prod^{2r-2}_{l=0}\left(1-\zeta^{r-l-1}e^{-2\pi n_a\frac{t^a}{\phi^0}}\right)^{(-1)^{r+l}\left(\substack{2r-2\\l}\right)d^r_{n_a}}~,
\end{equation}
and $\overline{Z}_{M2}$ for the anti-M2 brane trace. Here $t^a$ denotes the complexified Kahler class of the holomorphic cycle, that is, $t^a=p^a+i\phi^a$ in the near horizon variables. The first product comes from M2 branes wrapping genus zero curves while the second corresponds to the wrapping of genus $r$ curves.  

If $M$ has the form $T^2\times K3$ then the counting simplifies considerably. Firstly the supergravity modes contribution \eqref{eq:zsugra} is trivial since we have $\chi(M)=0$. On the other hand the M2 brane contribution is independent of the parameter $\zeta$ and gives simply
\begin{equation}\label{ZM2}
 Z_{M2}=\prod_{n_1>0}^{\infty}(1-e^{-2\pi n_1\frac{t^1}{\phi^0}})^{-\chi(K3)}~, \quad \chi(K3)=24~,
\end{equation}
 where, in this case, $t^1$ is the Kahler parameter of $T^2$.  For $\chi_{10}$, the gravity counting \eqref{ZM2} is precisely the infinite product in \eqref{res form}, with a suitable identification of the supergravity variable $t^1/\phi^0=(p^1+i\phi^1)/\phi^0$ with $\tau_2+i\tau_1$.

Both \eqref{ZM2} and \eqref{res form} are valid in a large central charge limit, i.e. $m\gg1$ and $\tau_{1,2}\sim \Lambda^0$. However we know that at finite central charge the counting of chiral primaries on $AdS_3$ suffers from a stringy exclusion principle \cite{Maldacena:1998bw} and the result of \cite{Gaiotto:2006ns} does not take this into account.  In addition the expressions \eqref{ZM2} and \eqref{res form} are not spectral flow invariant. The large central charge limit allows us to relax both the stringy exclusion principle and the spectral flow symmetry, and thus sum over all the chiral primaries in supergravity with no restriction. It is interesting to note that for $\chi_{10}$ one can take into account the exclusion principle in \cite{Maldacena:1998bw} by studying instead \eqref{eq:res}.  In \cite{Gomes:2015xcf,Murthy:2015zzy} the result for the microscopic degeneracy is shown to be a finite sum of Bessel functions, which is suggestive of a Rademacher expansion. As a matter of fact, the factors that multiply the Bessel functions can be identified with the polar coefficients \cite{Murthy:2015zzy} in a mock-Jacobi expansion \cite{Dabholkar:2012nd}.  Since under spectral flow from R to NS sector, the polar state degeneracy is mapped to the counting of chiral primaries, one can use the residue formula to make interesting predictions about the spectrum of KK fields of the dual bulk theory. Moreover, on a general note the study of the polar coefficients gives non-trivial information about both perturbative and non-perturbative corrections to black hole entropy. It would be interesting to address these corrections using localization techniques as in \cite{Gomes:2015xcf,Dabholkar:2010uh,Dabholkar:2011ec}.

The general lesson that we draw from $\chi_{10}$ is that the residue can be interpreted as counting fluctuations around the black hole, and these fluctuations are governed by the gravitational modes in question: basically counting the chiral primaries in the supergravity spectrum. Since \eqref{eq:res} and \eqref{res form} is as well applicable for other SMFs, it is rather natural to speculate that we can give a gravitational interpretation for our examples in section \ref{sec:CY} that involve elliptic genera of other Calabi-Yau manifolds.  The task we have ahead of us is to craft the appropriate supergravity theory which has the right spectrum to account for all the factors in \eqref{res form}.\footnote{Note that if we consider a different Calabi-Yau manifold in  \eqref{eq:zm2} and \eqref{ZM2}, we will not obtain \eqref{res form}. This makes it more challenging to decode which gravitational theory is the dual to the examples listed in section \ref{sec:CY}.} We leave the answers to these questions for future work.

\section{Discussion}\label{sec:discussion}

The goal of this work was to generalize the black hole microstate counting from the well-known examples such as $\chi_{10}$ to more general setups. We have shown that many of the methods used in the original setup can also be applied for general Siegel modular forms. More importantly, we also found novel examples of  SMFs that satisfy the criteria in section \ref{sec:msp}: SMFs  that have the correct features to account for black hole entropy, and seemingly a supergravity regime, includes the  exponential lifts of Elliptic Genera of Calabi-Yau manifolds in section \ref{sec:CY}, and the combination of $\chi_{12}$ and $\chi_{10}$ in section \ref{sec:x12}.  This shows how we can easily build and characterize SMFs that have the potential to describe black holes or other gravitational systems. 

We found that the leading contributions and the logarithmic corrections only depend on some very elementary data, namely the weight of the form, the poles and their residues; and we have cast this data also in terms of the Jacobi form that defines the exponential lift of the SMF. This gives the logarithmic correction in \eqref{eq:BHfull} a microscopic interpretation in terms of CFT data. In this sense the entropy is quite universal. For forms with the same type of poles as $1/\chi_{10}$, that is with divisors given by $H_1(1)$ listed in section \ref{sec:CY}, we were able to repeat the original arguments fully and thus give complete results. For more complicated divisors, it is clear that the same methods still work, but doing the analysis will require some more work. We were able to obtain the entropy in some of the regimes, but our analysis for $\chi_{35}$ shows that the results can be different, such as having a Hagedorn regime rather than Cardy regime. The main technical issues here are obtaining efficient expressions for the residue, and choosing the contours of integration carefully. It will be very interesting to explore the behavior of $d(\Qc)$ for more general Humbert surfaces. A first direction along this lines is when the Humbert surface is $H_D(0)$, i.e. extending the analysis of $\chi_{35}$ for other SMFs.

This is related to a second point. In this paper we only considered the growth of black hole states, that is states with polarity $\Qc\cdot\Qc > 0$. To investigate potential gravity duals, knowing the perturbative spectrum, \ie the polar states, is just as important. The growth of these states will for instance help to decide whether we are looking at a supergravity theory, or a full blown string theory. Moreover the growth will be related to how far the Cardy regime extends. Our results for $\chi_{35}$ for instance indicate that the polar states grow like Hagedorn.
Using similar techniques as in \cite{Benjamin:2015vkc}, one can indeed check this explicitly. This is in contrast to $\chi_{10}$, where the polar states agree with just the supergravity spectrum, which of course grows much more slowly. We expect that the examples in section \ref{sec:CY} can also be identified with a supergravity theory, which we will settle in future work. 

Ultimately of course the goal is to decide if our SMF have gravity duals. For this we still need to better understand the spectrum, as discussed in section \ref{sec:gravint}. It is particularly important to understand in what cases we have supergravity growth and a Cardy regime that extends even to $E\ll c$, and in which cases we have Hagedorn growth. In the former case, it may be possible to construct corresponding supergravity solutions. In the latter case, we should not expect such supergravity solutions, since the duals will be intrinsically stringy. Moreover the appearance of nearly holomorphic Jacobi forms suggests that some of the duals will have less supersymmetry than what we are used to. We leave this for future work.

In the context of finding counting formulas with black hole features, it would be interesting to explore if we can relax the condition of having $Sp(4,\Z)$ as a symmetry. This exchange symmetry among $\rho$ and $\tau$ is usually interpreted in the gravitational theory as electro-magnetic duality. This is a symmetry we know does not persist at the full quantum level for many black holes (such as ${\cal N}=2$ theories in 4D and 5D solutions). It would be interesting to investigated if we can achieve an extended Cardy regime by exploiting a different symmetry of the generating function. Along these lines, an interesting example is to understand in this language the counting formula for BPS states in ${\cal N}=8$ supergravity: there are exact formulas for the index \cite{MaldacenaMooreStrominger1999,ShihStromingerYin2006}, and it would be interesting to investigate if there are related forms with similar mathematical properties.

Finally, we would like to mention wall crossing phenomena. They are intimately related to the meromophicity of the generating functional, leading to interesting properties related to Mock modularity \cite{Dabholkar:2012nd}. On the gravitational side the interpretation of wall crossing is in terms of multicentered black holes \cite{ChengVerlinde2007,Sen:2008ta}. In contrast to $\chi_{10}$, where the order of the pole is $m_{1,1}=2$, our other examples have generically higher order poles, a phenomenon that also appeared in a different context in \cite{Huang:2015sta}. This may have interesting repercussions for the phase space of black holes. 

\section*{Acknowledgements}

It is a pleasure to thank Kathrin Bringmann, Miranda Cheng, Francesca Ferrari, Gerard van der Geer, Finn Larsen, Sameer Murthy and 
Ashoke Sen for helpful discussions.
CAK thanks the Harvard University High Energy Theory
Group for hospitality.   AB is supported by the Foundation for Fundamental Research on Matter (FOM). 
AC is supported by Nederlandse Organisatie voor Wetenschappelijk Onderzoek (NWO) via a Vidi grant. 
CAK is supported by the 
Swiss National Science Foundation through the NCCR SwissMAP. The work of AB, AC and JG is part of the Delta ITP consortium, a program of the NWO that is funded by the Dutch Ministry of Education, Culture and Science (OCW).

\appendix

\section{Jacobi forms}\label{app:jf}
\subsection{Properties of Jacobi forms}
 A Jacobi form $\varphi_{k,m}(\tau,z)$ is a holomorphic function on $\H\times\CC\rightarrow\CC$ that has the following defining properties:
 first, under modular transformations
\be\label{eq:jf1}
\varphi_{k,m}\le({a\tau+b\over c\tau +d},{z\over c\tau +d}\ri)= (c\tau +d)^k\exp\le({2\pi i m c z^2\over c\tau +d}\ri)\varphi_{k,m}(\tau,z)~,\quad  \forall \twobytwo{a}{b}{c}{d} \in SL(2,\ZZ)~,
\ee
and second, under translations
\be\label{eq:jf2}
\varphi_{k,m}\le(\tau,{z+ \lambda \tau +\mu}\ri)= \exp\le(-{2\pi i m (\lambda^2\tau+2\lambda z +\mu)}\ri)\varphi_{k,m}(\tau,z)~, \quad \lambda,\mu \in \ZZ~,
\ee
and it has Fourier expansion
\be\label{eq:jf3}
\varphi_{k,m}(\tau,z)= \sum_{\substack{n\geq0,l\\ 4mn\geq l^2}} c(n,l) q^n y^l~,\qquad q = e^{2\pi i\tau}\ , \qquad y = e^{2\pi i z}\ .
\ee
We define the \emph{discriminant} $\Delta:= 4nm-l^2$.
The coefficients $c(n,l)$ then only depend on $\Delta$ and $l$ (mod $2m$),
and in fact only on $\Delta$ if $m$ is prime. We will denote
the space of Jacobi forms of weight $k$ and index $m$ by
$J_{k,m}$.  

There are several special cases and generalizations 
of Jacobi forms which have to do with the summation range
in (\ref{eq:jf3}). \emph{Jacobi cusp forms} are Jacobi forms
for which $c(0,l)=0$. In particular they vanish at the cusp
$\tau = i\infty$.

\emph{Weak Jacobi forms} are holomorphic functions that
satisfy (\ref{eq:jf1}) and (\ref{eq:jf2}), 
but for which we don't impose the condition that $c(n,l)=0$ if 
$\Delta<0$. One can however show that we have $c(n,l)=0$ if
$\Delta < - m^2$, leading to a Fourier expansion
\be
\varphi_{k,m}(\tau,z)= \sum_{\substack{n\geq0,l\\ 4mn- l^2\geq -m^2}} c(n,l) q^n y^l\ .
\ee
Note that we are still only summing over $n$.

\emph{Nearly holomorphic Jacobi forms} finally satisfy
(\ref{eq:jf1}) and (\ref{eq:jf2}), but are allowed to have a
pole at the cusp $q=0$. More precisely, $\varphi$ is a nearly
holomorphic Jacobi form if there is a non-negative $n$ such
that $\Delta(q)^n \varphi$ is a Jacobi form.
In total we thus have the inclusions
\be
J^{cusp} \subset J \subset J^{weak} \subset J^{nh}\ .
\ee
We will mostly work with weak Jacobi forms.
A few commonly used weak Jacobi forms are
\bea\label{eq:jfex}
\phi_{10,1}(\tau,z)&=&\eta(\tau)^{18}\theta_1(\tau,z)^2~,\cr
&&\cr
\phi_{-2,1}(\tau,z)&=& {\theta_1(\tau,z)^2\over \eta(\tau)^6}~,\cr
&&\cr
\phi_{0,1}(\tau,z)&=& 4\le({\theta_2(\tau,z)^2\over \theta_2(\tau)^2}+{\theta_3(\tau,z)^2\over \theta_3(\tau)^2}+{\theta_4(\tau,z)^2\over \theta_4(\tau)^2}\ri)~,\cr
&&\cr
\phi_{-1,2}(\tau,z)&=& {\theta_1(\tau,2z)^2\over \eta(\tau)^3}~,
\eea
where $\eta(\tau)$ and $\theta_i(\tau,z)$ are the usual Dedekind eta function and theta functions, and $\theta_i(\tau)\equiv\theta_i(\tau,0)$. In some portions of the text we use $\Delta(\tau)\equiv \eta(\tau)^{24}=\frac{1}{1728}(E_4^3-E_6^2)$.
In fact, the ring of weak Jacobi forms of integer index is freely generated, 
\be
J^{weak} = \CC[E_4,E_6,\phi_{0,1},\phi_{-2,1},\phi_{-1,2}]/(432\xFive^2-\xThree \xFour^3+3\xOne \xThree^3\xFour-2\xTwo \xThree^4)\  .
\ee 
The space $J^{weak}_{k,m}$ of weak Jacobi forms of weight $k$ and
index $m$ is given by appropriate polynomials of the generators.
The space of half-integer index weak Jacobi form is closely
related to $J^{weak}_{k,m}$. We have
\be
J^{weak}_{2k,m+\frac12}= \phi_{0,\frac{3}{2}}J^{weak}_{2k,m-1}\ , \qquad
J^{weak}_{2k+1,m+\frac12}=\phi_{-1,\frac12}J^{weak}_{2k+2,m}\ ,
\ee
where 
\bea\label{eq:exwj}
\phi_{0,\frac{3}{2}}(\tau,z)=\frac{\theta_1(\tau,2z)}{\theta_1(\tau,z)}
&=&y^{-1/2}\prod_{n=1}^\infty(1+q^{n-1}y)(1+q^n y^{-1})(1-q^{2n-1}y^2)(1-q^{2n-1}y^{-2})\cr
\phi_{-1,\frac12}(\tau,z)=\frac{\theta_1(\tau,z)}{\eta(\tau)^3}
&=& -y^{-1/2} \prod_{n=1}^\infty (1-q^{n-1}y)(1-q^ny^{-1})(1-q^n)^{-2}
\eea 
 Another weak Jacobi form used in the main text is 
 \bea\label{eq:exwj1}
 \phi_{0,2}(\tau,z)&=&{1\over 2}\eta(\tau)^{-4}\sum_{m,n\in \Z}(3m-n)\le({-4\over m}\ri)\le({12\over n}\ri) q^{{3m^2+n^2\over 24}}y^{{m+n\over 2}}\cr
 &=& (y+4+y^{-1})+q(y^{\pm3}-8y^{\pm2}-y^{\pm1}+16)+q^2(\cdots)~.
 \eea

\subsection{Jacobi forms as partition functions}\label{sec:JFCFT}
Jacobi forms appear in physics quite often in the form of
generalized partition functions. Consider a fermionic CFT with
a $U(1)$ current $J$. In the Ramond sector
partition function with the fermion parity operator 
$(-1)^F$ inserted, $\tau$ and $z$ are then the chemical
potentials of the weight and $U(1)$ charge respectively.
(\ref{eq:jf1}) is then a direct consequence of the modular
transformation properties of genus 1 amplitudes.
(\ref{eq:jf2}) is usually interpreted as invariance
under spectral flow by one unit.

The most common setup for this is the elliptic genus
of theories with at least $N=(2,2)$ superconformal symmetry.
In that case we can define the elliptic genus as
\be
\chi(\tau,z)= \Tr_{RR}(-1)^F(-1)^{\bar F}q^{L_0-c/24}{\bar q}^{\bar L_0 -c/24}y^{J_0}\ ,
\ee 
which due to right-moving supersymmetry is indeed independent
of $\bar q$. Left-moving supersymmetry gives the constraint
\be
L_0 - \frac{c}{24} = (G_0)^2 \geq 0\ ,
\ee
which means that in (\ref{eq:jf3}) $n\geq 0$. It follows
that $\chi$ is a weak Jacobi form, in fact of weight 0
and index $m = c/6$.
A typical example of this is the non-linear sigma model
on K3, whose elliptic genus is simply 
\be
\chi_{K3}(\tau,z)=2\phi_{0,1}(\tau,z)\ .
\ee
For Calabi-Yau threefolds, there is a one-dimensional vector space.
The elliptic genus depends only on the Hodge numbers and
is given by
\be
\chi(\tau,z)=(h^{1,1}-h^{2,1})\phi_{0,3/2}(\tau,z) = \frac12 e(M) \phi_{0,3/2}(\tau,z)\ .
\ee

Note that generically we expect the elliptic genus to be
a weak Jacobi form, and not a Jacobi form. The coefficients
of Jacobi forms grow only polynomially, whereas we expect the
underlying CFT to have Cardy growth, which agrees with the
exponential growth of weak Jacobi forms. In this supersymmetric
setup we can in particular spectrally flow by half a unit from
the NS sector to the Ramond sector and vice versa. The left-moving
NS vacuum is then mapped to the term $y^{c/6}$, which is clearly
a polar state, so that the resulting form is a weak Jacobi form.
It is however possible that due to an enhanced symmetry, the
Witten index of the right-moving  states that couple to the left-moving
vacuum vanishes, so that this term does not appear in the elliptic
genus, and similar for all other polar terms. Unless such 
cancellations happen however, we will get a weak Jacobi form and not a Jacobi form.

On the other hand we can try to generalize the elliptic
genus to CFTs with less supersymmetry. To ensure that the
resulting partition function is (nearly) holomorphic,
we need either a chiral (purely left-moving) theory,
or at least $N=1$ supersymmetry for the right-movers,
so that we can repeat the Witten index argument. On the
left-moving side we need a $U(1)$ current. (\ref{eq:jf1})
is then automatically satisfied, see \eg \cite{Kraus:2006wn,Benjamin:2016fhe}
and references therein. To satisfy (\ref{eq:jf2}),
it is enough to have something similar to integer
spectral flow. A sufficient condition for that
is to only have states of integer $U(1)$ charge: The idea
is that we can bosonize the $U(1)$ current, and under
suitable integrality conditions on the charges,
$e^{i\phi}$ is then a local operator that induces
integral spectral flow. In such a theory we no longer
have the condition that $n\geq0$, which would allow for
nearly holomorphic Jacobi forms as partition functions.

 More generally, $\varphi_{k,m}(\tau,z)$ are also the building blocks of for chiral CFTs, as well as warped CFTs \cite{Detournay:2012pc,Castro:2015uaa}. And in particular, many of the properties of exponential lifts discussed here in principle apply to these non-supersymmetric theories. For example, it is simple to construct a SMF for symmetric products of the Monster CFT in \cite{Witten:2007kt}. But unfortunately other examples not involving supersymmetry are much less developed.  

\subsection{Hecke operators}\label{app:Hecke}
\cite{9781468491647} defines three types of Hecke operators
acting on $\phi = \sum_{n,r}c(n,r)q^ny^l$:
\bea\label{eq:hecke}
U_p &:& J_{k,m} \rightarrow J_{k,mp^2}\cr
V_p &:& J_{k,m} \rightarrow J_{k,mp}\cr
T_p &:& J_{k,m} \rightarrow J_{k,m} 
\eea
Note that they all map Jacobi forms to Jacobi forms.
$U_p$ simply maps $\phi(\tau,z) \mapsto \phi(\tau,pz)$,
so that
\be
\phi|U_p = \sum_{n,l} c(n,l/p)q^n y^l\ .
\ee
(Our convention is to take $c(n,l)=0$ if its arguments
are non-integer or outside the original summation range.)
From this it is clear that it also maps weak Jacobi
forms to weak Jacobi forms. In \cite{MR1616929} this
is essentially the operator $\Lambda_p$.

$V_l$ is probably the best known Hecke operator.
It acts as
\be
\phi|V_p = \sum_{n,l} \sum_{a|(n,l,p)}a^{k-1}c\left(\frac{np}{a^2},\frac{l}{a}\right)q^ny^l\ .
\ee
It is again clear that this also maps
weak Jacobi forms to weak Jacobi forms.
Physically it is related to cyclic orbifolds, and
therefore shows up crucially in symmetric orbifold
partition functions. In \cite{MR1616929} this
is essentially the operator $T_-(p)$.

$T_p$ finally is the operator in \cite{MR1616929}
denoted by $T_0(p)$. For $\phi$ with weight $k=0$
it gives a new form with coefficients
\be
c_p(n,l)=p^3c(p^2n,pl)+G_p(n,l,m)c(n,l)+ \sum_{\lambda\mod p}
c\left(\frac{n+\lambda l+\lambda^2 m}{p^2},\frac{l+2\lambda m}{p}\right)
\ee 
where the Gauss sum is
\be
G_p(n,l,m)=-p+\sum_{a,b\mod p} \exp\left(2\pi i\frac{na+lab+mab^2}{p}\right)\ .
\ee
Very importantly, $T_p$ does \emph{not} map weak Jacobi forms
to weak Jacobi forms: in general it will give nearly holomorphic
Jacobi forms instead.

\section{Logarithms in the Cardy regime: $n\gg m$}\label{app:cardy}

In this appendix we derive the logarithmic correction to the leading asymptotic growth of a Jacobi form in the Cardy regime; this is the regime that we denote {\bf I} in section \ref{sec:asymp}. Consider the follow integral
\be\label{eq:c1}
d(n,l)= \int d\tau dz\, e^{-2\pi i  \tau n -2\pi i  z l} \varphi_{k,m}(\tau,z)~,
\ee
where $\varphi_{k,m}(\tau, z)$ is a weak Jacobi form of weight $k$ and level $m$. 
%
%
%
We want to estimate $d(n,l)$ for large values of $n$ and $l$, and the easiest is to do a saddle point approximation. First we rewrite \eqref{eq:c1} as
\be\label{eq:c2}
d(n,l)= \int d\tau dz\, {1\over \tau^k}  \exp\le(-2\pi i n \tau -2\pi i l z-{2\pi i m  z^2\over \tau }\ri) \varphi_{k,m}\le(-{1\over\tau},{z\over \tau}\ri)~,
\ee
where we used \eqref{eq:jf1}. Assuming that the saddle is at some small imaginary value of $\tau$ and $z$ of order one,\footnote{One can weaken the assumption that $\tau$ is small, but this requires further restrictions on the spectrum of $\varphi(\tau,z)$ which are not needed for the purposes of this section. See e.g. \cite{Hartman:2014oaa} for a recent discussion on such restrictions.} we can approximate $\varphi_{k,m}\le(-{1\over\tau},{z\over \tau}\ri)$ by its most polar term
\be\label{eq:polar1}
\varphi_{k,m}\le(-{1\over\tau},{z\over \tau}\ri)\sim \exp(-{2\pi i n_0\over \tau})  \exp({2\pi i l_0 z \over \tau})~,
\ee
where we assigned to the most polar state charges $(n,l)=(n_0,l_0)$, i.e. the state with most negative discriminant $\Delta$.\footnote{There could be a degeneracy associated to the most polar state, but it is negligible within the approximation taken here.} This gives at leading order 
\be
d(n,l)\sim \int d\tau dz\, {1\over \tau^k}  \exp\le(-2\pi i n \tau -2\pi i l z-{2\pi i m  z^2\over \tau }\ri)\exp(-{2\pi i n_0\over \tau})  \exp({2\pi i l_0 z \over \tau})~.
\ee
The saddle point of the above integral is at 
\bea
\tau_\star&=& \sqrt{n_0- l_0^2/4m\over n-l^2/4m}~,\cr
z_\star&=& {l_0\over 2m}- {l\tau_\star\over2 m}~.
\eea
We assumed that the original integral was dominated by the behavior of $\varphi_{k,m}$ at $\tau\to i0^+$ and $z\sim O(1)$: this requires that $n-l^2/4m \gg |n_0 - l_0^2/4m|$ and $n_0-l_0^2/4m<0$. For many systems, the discriminant of the most polar term is related to the central charge $c$ of the system, i.e. $n_0 - l_0^2/4m \sim c$, hence we are estimating $d(n,l)$ for $n\gg c$. 
Performing the saddle point approximation gives 
\bea\label{eq:c4}
d(n,l) &\sim& {\tau^{2-k}_\star}  \exp(-{4\pi i \over \tau_\star}(n_0- {l_0^2\over4m}))\cr
&\sim&  {\tau^{2-k}_\star}  \exp\le(2\pi \sqrt { (-4n_0 +{l_0^2\over m}) (n-{l^2\over 4m}) } \ri)~.
\eea
The leading exponential contribution is the usual Cardy formula; the polynomial contribution controlled only by the weight of $\varphi_{k,m}$. These corrections should be compared with the analogous logarithmic correction in other scaling regimes.

\section{Black hole near-horizon geometry and attractor equations}\label{app:BH}

In this appendix we revisit the D1-D5-P-KK system in IIB string theory on $K3\times S^1\times \tilde{S}^1$. This configuration preserves four out of sixteen supercharges of four dimensional $\mathcal{N}=4$ supergravity. We describe the brane Kaluza-Klein monopole configuration and the black hole near-horizon geometry that it gives rise to. As we show, it is easier to analyze the geometry from the IIA point of view where the attractor equations are naturally embedded in $\mathcal{N}=2$ supergravity. We will map first the configuration to IIA and then to M-theory where it has an interpretation as a MSW string. 

This configuration consists of $Q_5$ D5-branes wrapping $K3\times S^1$, $Q_1$ D1-branes wrapping $S^1$, $K$  Kaluza-Klein monopoles associated with the circle $\tilde{S}^1$ and $n$ units of momentum along the circle $S^1$. For $K=1$ this is equivalent to putting the D1-D5-P system on the background of a Kaluza-Klein monopole. Since the Taub-Nut geometry approaches $\mathbb{R}^3\times \tilde{S}^1$ at infinity while it is $\mathbb{R}^4$ near the origin, this provides a five/four dimensional connection, which relates the four dimensional black hole to the BMPV black hole. Following \cite{Dabholkar:2010rm}, we start by performing a mirror symmetry transformation on $K3$, which take us to a D3-brane configuration. We have now $Q_1$ D3-branes wrapping $\gamma\times S^1$ and $Q_5$ D3-branes wrapping $\tilde{\gamma}\times S^1$ where $\gamma,\tilde{\gamma}$ are a pair of dual 2-cycles in $K3$. A T-duality along the circle $\tilde{S}^1$ takes the D3 to D4 branes wrapping $\gamma\times S^1\times \hat{S}^1$ and similarly for the dual cycle, with $\hat{S}^1$ the T-dual circle. The Kaluza-Klein monopoles map to $K$ NS5-branes wrapping $K3\times S^1$. From the M-theory point of view we have $Q_1$ M5-branes wrapping $\gamma\times S^1\times \hat{S}^1\times S^1_M$, $Q_5$ M5-branes wrapping $\tilde{\gamma}\times S^1\times \hat{S}^1\times S^1_M$, K M5-branes on $K3\times S^1\times \hat{S}^1\times S^1_M$ and $n$ units of momentum along the circle $S^1$ where $S^1_M$ is the M-theory circle.  From the eleven dimensional point of view we can reduce instead along the circle $S^1$. For convenience denote the circle $S^1$ by $\bar{S}^1_M$ so we know that we are reducing M-theory along this circle. The final configuration corresponds  to a MSW string configuration consisting of an M5-brane wrapping the cycle $P\times \bar{S}^1_M$ where $P$ is a four cycle in the class $P=p^a\Sigma_a\in H_4(K3\times \hat{S}^1\times S^1_M)$ and $\Sigma=(\gamma\times \hat{S}^1\times S^1_M,\tilde{\gamma}\times\hat{S}^1\times S^1_M,K3)$. In addition one has $n$ units of momentum along the circle $\bar{S}^1_M$. From the type IIA point of view, this configuration gives rise to a D4-D0 black hole. We have $p^a$ D4-branes wrapping cycles $\Sigma_a$ and $n$ D0-branes. To introduce D2-branes we can start form the IIB configuration and consider momentum along the circle $\tilde{S}^1$. Under the chain of dualities described above this gives rise to D2-branes wrapping 2-cycles dual to $\Sigma_a$.   

Summarizing one has the IIA D-brane configuration
\begin{eqnarray}
&&D4:\;p^a\Sigma_a,\;\Sigma_a\in H_4(\mathbb{Z})\\
&&D2:\;q_a\tilde{\Sigma}^a,\; \tilde{\Sigma}^a\in H_2(\mathbb{Z})\\
&&D0:\;q_0
\end{eqnarray}
For example, for the D1-D5-P-KK configuration one has $\Sigma_1=K3$ and $\Sigma_{a\neq 1}=\gamma_a\times T^2$ with $\gamma_a\in H_2(K3,\mathbb{Z})$. Similarly $\tilde{\Sigma}^1=T^2$ and $\tilde{\Sigma}^{a\neq 1}=\tilde{\gamma}_a$. 

It is also useful to map the D1-D5-P-KK system to a configuration in Heterotic string theory compactified on $T^6$, consisting only of NS-NS charges. This theory has a U-duality group $G$ consisting of the following factors
\begin{equation}
G(\mathbb{Z})=SL(2,\mathbb{Z})\times SO(6,22;\mathbb{Z})~.
\end{equation}
The $SL(2,\mathbb{Z})$ factor is responsible for electric magnetic duality while the second factor corresponds to the T-duality group. A dyon charge configuration transforms in the fundamental of $SL(2,\mathbb{Z})$ and in the vector representation of $SO(6,22;\mathbb{Z})$.  Lets denote the dyon configuration by the  vector
\begin{equation}
\Gamma=\left(\begin{array}{c}
Q^i\\ P^i
\end{array}\right),\;i=1\ldots 28~.
\end{equation}
Here $Q^i$ and $P^i$ are respectively the electric and magnetic charge vectors. The subscript $i$ denotes that it transforms under the vector representation of  $SO(6,22;\mathbb{Z})$. In addition, the  dyon $\Gamma$ transforms under electric-magnetic duality as 
\begin{equation}
\left(\begin{array}{c}
{Q'}^i\\ {P'}^i
\end{array}\right)=\left(\begin{array}{cc}
a & b\\
c & d
\end{array}
\right)\left(\begin{array}{c}
Q^i\\ P^i
\end{array}\right),\;\;\left(\begin{array}{cc}
a & b\\
c & d
\end{array}
\right)\in SL(2,\mathbb{Z})~.
\end{equation}
It is natural to associate to the dyon the following T-duality invariant combinations
\begin{equation}\label{eq:Tcharges}
Q^2\equiv Q^iL_{ij}Q^j,\quad P^2=P^iL_{ij}P^j,\quad Q\cdot P\equiv Q^iL_{ij}P^i ~,
\end{equation}
where $L_{ij}$ is the $SO(6,22)$ invariant metric. Furthermore, it is easy to check that the triplet $(Q^2,P^2,Q.P)$ transforms under the vector representation of $SO(1,2)$ or the symmetric representation $\textbf{3}$ of $SL(2,\mathbb{Z})$. 

Under the map from Heterotic on $T^4\times S^1\times \tilde{S}^1$ to IIA on $K^3\times S^1 \times \tilde{S}^1$, the charges are assigned in the following way
\begin{eqnarray}
Q^i&=&(q_0,p^1,q_a)~,\cr
P^i&=&(-q_1,p^0,p^a)~.
\end{eqnarray} 
We have not included the charges corresponding to momentum/winding and NS-5/KK monopole associated with the circles on $T^4$.
For example $q_0$ D0-branes map to momentum along the circle $S^1$ while $p^1$ D4-branes wrapping $K^3$ map to $p^1$ units of winding along the circle $S^1$. The T-duality invariants are therefore
\begin{eqnarray}
Q^2&=&2q_0 p^1+D^{ab}q_aq_b~,\cr
P^2&=&-2q_1p^0+D_{ab}p^ap^b~,\cr
Q\cdot P&=&p^0q_0-p^1q_1+p^aq_a~,
\end{eqnarray}
where $D_{ab}$ is the intersection matrix of $K3$, that is, $D_{ab}=\int \gamma_a\wedge \gamma_b $.

We now describe the near-horizon geometry of the four dimensional D4-D2-D0 black hole. The analysis is valid for any Calabi-Yau manifold and relies only on the four dimensional $\mathcal{N}=2$ supergravity. Further details of the attractor equations can be found in \cite{Ferrara:1995ih}. 

The ten dimensional near-horizon metric is 
\begin{equation}
ds_{10}^2=\frac{L^2}{4}\left(-(r^2-1)dt^2+\frac{dr^2}{r^2-1}+d\theta^2+\sin^2(\theta)d\phi^2\right)+ds^2_{\rm CY}~.
\end{equation}
The graviphoton field is given by
\begin{equation}
F^0=\phi^0dr\wedge dt~,
\end{equation}
while the remaining $n_V$ vectormultiplet gauge fields, that couple to the D2 charges, have field strength
\begin{equation}\label{vector gauge field}
F^a=-\phi^adr\wedge dt+p^a\sin(\theta)d\theta\wedge d\phi~,
\end{equation}
with $p^a$  the magnetic charges. The scalar fields, on the other hand, are determined in terms of the electric fields and the magnetic charges, that is,
\begin{equation}\label{attractor 1}
LX^0=\phi^0,\quad LX^a=\phi^a+ip^a~,
\end{equation}
with $p^0=0$.
The remaining hypermultiplet scalar fields are freely adjustable parameters throughout the black hole geometry. Finally the Kahler form of the Calabi-Yau is determined by
\begin{eqnarray}
J\propto \frac{p^a}{L}~.
\end{eqnarray}

The M-theory lift of this configuration corresponds to promoting the graviphoton gauge field as the Kaluza-Klein gauge field associated with the M-theory circle. The uplifted geometry contains now a local $AdS_3$ factor,
\begin{equation}\label{near horizon geometry}
ds^2_{11}=\frac{L^2}{4}\left[ds^2_{AdS_2}+\frac{1}{(\phi^0)^2}\left(dy-\phi^0(r-1)dt\right)^2+d\Omega^2\right]+ds^2_{\rm CY}~,
\end{equation} 
where $ds^2_{AdS_2}$ and $d\Omega^2$ denote unit size metrics for $AdS_2$ and $S^2$ respectively, and $y$ is the M-theory circle. The vector-multiplet gauge fields uplift to
\begin{eqnarray}
A^a_{5D}=-\frac{\phi^a}{\phi^0}dy+A^a_{\text{Dirac}}~,
\end{eqnarray}
with $A^a_{\text{Dirac}}$ the Dirac monopole gauge field that gives rise to the magnetic flux (\ref{vector gauge field}).

The attractor equations relate the values of the scalar fields to the charges of the black hole in the following way
\begin{equation}\label{attractor 2}
\text{Im}(L \mathcal{F}_I)=q_I~,\quad I=0\ldots n_V~,
\end{equation}
where $\mathcal{F}_I=\partial_{X^I}\mathcal{F}(X)$ with $\mathcal{F}(x)$ the  prepotential of $\mathcal{N}=2$ supergravity. The attractor equations (\ref{attractor 2}) arise from the extremization of the functional
\begin{equation}\label{eq:EBH}
S(\phi,p,q)=-\pi q_I\phi^I+\pi \text{Im}\mathcal{F}(\phi+i p)~.
\end{equation}
This fact has led to the conjecture that the black hole partition function equals the square of the topological string partition function $Z_{BH}=|Z_{\text{top}}|^2$ \cite{Ooguri:2004zv}. Recently, it was shown that $S(\phi,p,q)$ follows from a localization computation  on $AdS_2\times S^2$ \cite{Dabholkar:2010uh}.

In the two derivative approximation the prepotential is a polynomial of the scalar fields. For the $K3\times T^2$ compactification one has
\begin{equation}\label{prepotential}
\mathcal{F}(X)=-\frac{1}{2}\frac{X^1}{X^0}\sum_{a,b=2}^{23}D_{ab}X^aX^b~.
\end{equation}
Note that $a,b$ run over the 2-cycles of $K3$. This leads to the following attractor solutions
\begin{eqnarray}
&&\phi^a=-q^a\frac{\phi^0}{p^1}+p^a\frac{\phi^1}{p^1}~,\cr
&&\phi^1=-\frac{Q\cdot P}{P^2}\phi^0~,\cr
&&\phi^0=\frac{p^1 P^2}{\sqrt{Q^2P^2-(Q\cdot P)^2}}~.
\end{eqnarray} 

For a more general Calabi-Yau compactification, the two derivative prepotential (\ref{prepotential}) is instead given by
\begin{equation}
\mathcal{F}(X)=-\frac{1}{6}\sum_{a,b,c=1}^{b_2}D_{abc}\frac{X^aX^bX^c}{X^0}~,
\end{equation}
where $D_{abc}$ is the intersection matrix and $b_2$ is the dimension of $H^{(1,1)}$, the Kahler class. The attractor equations 
are modified as
\begin{eqnarray}
&&\phi^a=-q^a\phi^0~,\cr
&&\phi^0=\sqrt{\frac{P^3/6}{\hat{q}_0}}~,
\end{eqnarray}
where $\hat{q}_0=q_0-D_{ab}q^aq^b/2$ is the spectral flow invariant charge combination, and we have defined $P^3=D_{abc}p^ap^bp^c$ and $D_{ab}=D_{abc}p^c$.

Note that the attractor equations are invariant under the scaling symmetry
\begin{equation}
\phi^I\rightarrow \lambda \phi^I,\quad q_I\rightarrow \lambda q_I,\quad p^I\rightarrow \lambda p^I,\quad I=0\ldots n_V~.
\end{equation}
When $\lambda\gg 1$ the M-theory circle radius $~1/\phi^0$, in units of the $AdS$ length $L$, becomes very small. This corresponds to the string-theory limit. In the other limit, that is when $q\gg 1$ and $p$ is kept fixed we have that $1/\phi^0$ is of order one or bigger and thus the theory is well described in M-theory.

The number of 1/4BPS states of D1-D5-P-KK system is describe by the reciprocal of $\chi_{10}$. The notation  used here mapped to the one used in section \ref{sec:chi10} is as follows. The T-duality invariants \eqref{eq:Tcharges} in terms of the charge vector \eqref{eq:Qdef} is
\be\label{eq:Qm}
Q^2=2n~,\quad  P^2=2m ~,\quad l=Q\cdot P~,
\ee
and hence $2\Qc^2= Q^2 P^2 -(Q\cdot P)^2$. The entropy \eqref{eq:EBH} for a two derivative supergravity prepotential  is $S_{\rm BH}=\pi \sqrt{2\Qc^2}$, and in ${\cal N}=4$ supergravity the logarithmic correction vanishes. On the microscopic side this corresponds to the results in \eqref{eq:cr1}. It is important to emphasize that the dictionary \eqref{eq:Qm} is what dictates the scaling regime for which we need to evaluate $d(\Qc)$; without this data there is room for ambiguity.

\bibliographystyle{JHEP}
\bibliography{ref}

\end{document}